\documentclass[10pt]{iopart}
\pdfoutput=1
\usepackage{graphicx}
\usepackage{bm}
\usepackage{subfigure}
\usepackage{color}
\usepackage{xr}
\usepackage{times}

\newcommand{\eqref}[1]{(\ref{#1})}

\begin{document}

\title[Dynamics of an engineered quantum phase transition in a coupled binary BEC]{Causality and defect formation in the dynamics of an engineered quantum phase transition in a coupled binary Bose-Einstein condensate}
\author{Jacopo Sabbatini$^1$, Wojciech H Zurek$^2$, Matthew J Davis$^3$}
\address{$^1$The University of Queensland, School of Mathematics and Physics, ARC Centre of Excellence for Engineered Quantum Systems, Queensland 4072, Australia}
\ead{sabbatini@physics.uq.edu.au}
\address{$^2$Theory Division, Los Alamos National Laboratory, Los Alamos, New Mexico 87545, USA}
\address{$^3$The University of Queensland, School of Mathematics and Physics, Queensland 4072, Australia}

\date{\today}

\begin{abstract}

Continuous phase transitions occur in a wide range of physical systems, and provide a context for the study of non-equilibrium dynamics and the formation of topological defects.  The Kibble-Zurek (KZ) mechanism predicts the scaling of the resulting density of defects as a function of the quench rate through a critical point, and this can provide an estimate of the critical exponents of a phase transition.
  In this work we extend our previous study of the miscible-immiscible phase transition of a binary Bose-Einstein condensate (BEC) composed of two hyperfine states in which the spin dynamics are confined to one dimension [J. Sabbatini \emph{et al.}, Phys. Rev. Lett. \textbf{107}, 230402 (2011)]. The transition is engineered by controlling a Hamiltonian quench of the coupling amplitude of the two hyperfine states, and results in the formation of a random pattern of spatial domains. Using the numerical truncated Wigner phase space method, we show that in a ring BEC the number of domains formed in the phase transitions scales as predicted by the KZ theory.  We also consider the same experiment performed with a harmonically trapped BEC, and investigate how the density inhomogeneity  modifies the dynamics of the phase transition and the KZ scaling law for the number of domains.  We then make use of the symmetry between inhomogeneous phase transitions in anisotropic systems,  and an inhomogeneous quench in a homogeneous system, to engineer coupling quenches that allow us to quantify several aspects of inhomogeneous phase transitions.  In particular, we quantify the effect of causality in the propagation of the phase transition front on the resulting formation of domain walls, and find indications that the density of defects is determined during the impulse to adiabatic transition after the crossing of the critical point.

\end{abstract}

\pacs{03.75.Mn, 03.75.Lm, 05.70.Fh}

\maketitle

\section{Introduction}

The study of non-equilibrium quantum systems is exemplified by systems undergoing a quantum phase transition, in which the ground state undergoes a macroscopic change of symmetry as a Hamiltonian parameter is changed through a critical value \cite{Sachdev:0E31sdLn}. These phase transitions often result in the formation of \emph{topological defects}, occurring when the topology of the degenerate ground states of the new phase is non-trivial \cite{Vilenkin:vw}. In this case, particular classes of excited states of the system, the topological defects, cannot be continuously transformed to the ground state and their existence is therefore protected by topological conservation laws.  We can potentially unveil some of the mysteries behind out-of-equilibrium systems by studying the mechanism of formation of topological defects. These are imprinted in the system during the ``turbulent'' era of a phase transition but survive its subsequent evolution. Like fossils, topological defects provide information about a period that cannot be easily modelled analytically. The ability to engineer a quantum phase transition and recover information from topological defects is an invaluable resource for the study of non-equilibrium systems~\cite{Dziarmaga2010a,Polkovnikov2011a}.

The Kibble-Zurek mechanism is a theory developed by Kibble \cite{Kibble:1976vm} and Zurek \cite{Zurek:1985wh} with the goal of predicting the scaling of the density of  topological defects formed with the rate at which the critical point of a phase transition is crossed. The theory is based on the assumption that every system undergoing a phase transition experiences a phase of impulsive behaviour where the topological defects are ``crystallised'' in the system. The breakdown of adiabaticity, due to the divergence of the response time  near a critical point, is a characteristic of every continuous phase transition and the theory offers a model for the formation of defects across a wide range of systems, from cosmology \cite{Cruz:2007kf} to condensed matter \cite{Chuang:1991uz,Hendry:1994vt,Ruutu:1996ww,Bauerle:1996tf,Dodd:1998tc,monacorivers2006,monacorivers2002,Chae:2012jr,Griffin:2012tf}.  While initially applied to classical phase transitions occurring at finite temperature \cite{Kibble:1976vm,Zurek:1985wh}, more recently the scenario has been formulated and successfully applied to quantum phase transitions at zero temperature \cite{Zurek:2005cu,Damski:2005cr,Dziarmaga:2005jq,Polkovnikov:2005gr,Damski:2007ek,Damski:2008wo,Saito:2007ta,Damski:2009dw}.

The development of isolated, yet highly tunable ultra-cold gas experiments have provided exciting opportunities for the study of quantum matter and quantum dynamics. Recent advances in the production and manipulation of Bose-Einstein condensates (BEC) offer the possibility of observing a variety of quantum phase transitions, from the study of the Ising model with ultra-cold gases \cite{Simon:2011hu} to the exploration of the Mott insulator to superfluid phase transition \cite{Chen:2011iu}. Experimentalists in degenerate quantum gases have access to a large set of tools like Feshbach resonances \cite{Ketterle:1998if}, arbitrary trapping \cite{Henderson:2009eo} and single-atom imaging \cite{Bakr:2010gd} that offer them the ability to engineer Hamiltonians and to control their parameters. Furthermore the production of the first optically trapped BEC \cite{StamperKurn:1998tp} led to the exploration of the richer physics of multi-component condensates. Our ability to accurately manipulate and prepare a wide range of states of condensates with internal degrees of freedom allowed the observation of spin changing dynamics \cite{Kronjaeger:2008jr} and gave us access to the phase diagram of spinor BECs \cite{Sadler:2006fw}. Multi-component BECs were also the key piece for the first observation of spin-orbit coupling in neutral atoms \cite{Lin:2011ba}.

\begin{figure}
\begin{center}
\includegraphics[width=0.9\textwidth]{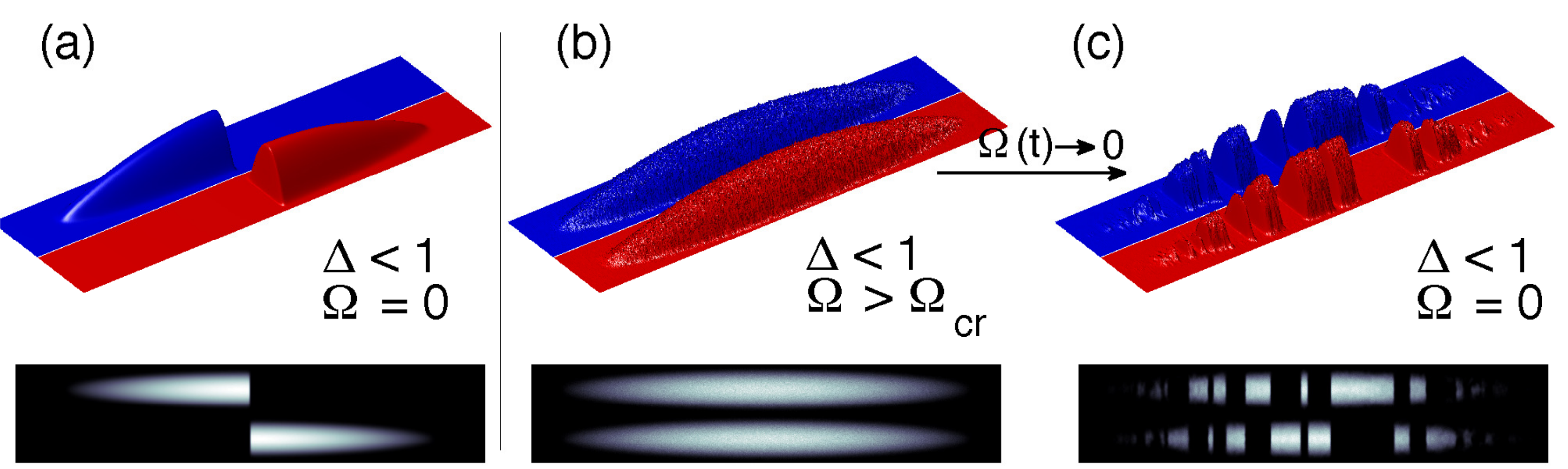}
\end{center}
\caption{Density of the two components of a binary BEC. (a) Natural ground state of a binary immiscible system. (b) In the strong coupling regime state (a) becomes miscible.  Quantum noise has been added that results in the formation of domains (see text). (c) Quenching the coupling $\Omega(t)$ to zero brings the system back to its immiscible phase. If the quench is non-adiabatic a random pattern of domains is created and the system is left in an excited state. We propose to test the Kibble-Zurek mechanism by counting the number of formed domains in function of the quench time.}\label{fig:SetUp}
\end{figure}

In this work we extend a proposal for an experiment to test the KZ theory using a binary BEC consisting of a two hyperfine states of a single atomic species~\cite{Sabbatini:2011gu}. Binary condensates can be classified as miscible or immiscible, depending on whether the two states can naturally coexist in space \cite{Ho:1996up}. However, it has been shown that introducing a coupling between the two states can transform an immiscible condensate into a miscible one, and that there is a quantum phase transition between the two states \cite{Merhasin:2005tv,Tommasini:2003tx}. In the strong coupling regime the mean-field ground state of the system consists of a superposition of the two states whose density then resembles the miscible phase.  Here we consider a naturally immiscible condensate beginning in the ground state of the strong coupling regime, which is subsequently driven back to the immiscible phase by quenching the coupling to zero (see Fig.~\ref{fig:SetUp}) \cite{Sabbatini:2011gu}.  If the quench is non-adiabatic, regions of the system that are not causally connected will undergo the phase transition independently and the final configuration is a random pattern of domains of the two components, Fig.~\ref{fig:SetUp}(c). The  KZ mechanism is demonstrated by counting the number of domains formed as a function of the characteristic quench time.  When applied to a BEC in a ring trap \cite{Henderson:2009eo,Gupta:2005ed,Ramanathan:2011bi} this scheme is an experimentally feasible candidate to provide a definitive demonstration of the KZ mechanism \cite{Sabbatini:2011gu}. In our study we perform our simulations using the truncated Wigner approximation (TWA) \cite{Blakie:2008vt} to numerically simulate the quantum dynamics of the BEC during the quench. This phase-space method maps quantum operators to stochastic variables \cite{Blakie:2008vt} to include quantum corrections to the mean-field dynamics, allowing us to capture the symmetry breaking that occurs in the phase transition.

In the vast majority of theoretical and experimental works, the KZ mechanism has been studied for homogeneous phase transitions \cite{Zurek:1985wh,Zurek:2005cu,monacorivers2002}, i.e.~phase transitions in which the critical point is traversed simultaneously for all spatial regions of the system.  However, ultra-cold quantum gases are often confined to trapping potentials which lead to spatially varying phases and inhomogeneous phase transitions \cite{Dziarmaga:2010dv}.  In such situations additional considerations, such as the effects of causality, can change the quantitative details of the phase transition, and the KZ theory needs to be modified \cite{Zurek:2009wx,delCampo:2011uv,Campo:2010ff}.  In the implementation we describe we can study the effects of inhomogeneities by engineering the phase transition in uniform systems.  Spatially shaping the coupling quench, for example, allows the possibility of ``simulating'' an inhomogeneous phase transition in systems with otherwise straightforward dynamics such as a homogeneous 1D ring BEC.  The precise level of control required over the phase transition can be experimentally achieved with the use of modern experimental techniques that, for example, offer the possibility of shaping the spatial intensity profile of a laser using holography \cite{Pasienski:2008hf,Gaunt:2011ut}. Engineered Hamiltonian quenches are potentially a powerful tool to study the characteristics of phase transitions in a controlled fashion that can be applied to theoretical and experimental studies of other quantum phase transitions.

This paper is structured as follows. In Sec.~\ref{sec:model} we introduce the model of binary condensates and outline the principles behind the KZ mechanism. In Sec.~\ref{sec:uniform} we model the miscible-immiscible phase transition in a ring BEC, and demonstrate that the number of defects scales with the quench time as predicted by the KZ theory~\cite{Sabbatini:2011gu}. We then consider the same phase transition in a harmonically trapped BEC in Sec.~\ref{sec:trappedBEC}. We find that the counting of domains can be problematic, and we find a different scaling exponent with respect to the uniform case.  In Sec.~\ref{sec:iptRing} we engineer a spatial quench of the coupling capable of ``simulating'' the phase transition of an harmonically trapped BEC in a ring geometry. This engineered phase transition allows us to verify the exponent observed in the previous section.  In Sec.~\ref{sec:rectifying} we extend the method of spatially shaping the coupling to invert the process, i.e. we attempt to simulate a \textit{homogeneous} phase transition in an \textit{harmonically trapped} system.  We quantify the effects of causality for an inhomogeneous phase transition by accurately identifying the subsonic and supersonic regimes of the phase transition in Sec.~\ref{sec:causality}. In Sec.~\ref{sec:beforeOrAfter} we demonstrate that for this system the domain walls are seeded after the passing of the critical point, and finally in Sec.~\ref{sec:feasibility}, we explore the experimental feasibility of our scheme and summarise our results.

\section{Model}\label{sec:model}

\subsection{Binary Bose-Einstein condensates}

Several authors have described the physics of  binary BEC mixtures \cite{Ho:1996up,Timmermans:1998tj} and coupled binary BECs \cite{Gligoric:2010bi,Lee:2009ib}.   We define the Hamiltonian for our coupled $1$D system as
\begin{equation}
\hat{H} =  \hat{H}_{0} + \hat{H}_{\rm I} + \hat{H}_{\rm C},\label{eq:fullHam}
\end{equation}
where $\hat{H}_{0}$, $\hat{H}_{\rm I}$, and $\hat{H}_{\rm C}$ are the single-particle, interaction, and coupling Hamiltonians respectively, defined as
\begin{eqnarray}
\hat{H}_{0} & =  & \int dx \sum_{i=1}^{2}\hat{\psi}_{i}^{\dagger}(x)\left[ -\frac{\hbar}{2m}\frac{\partial^2}{\partial x^2}+V(x)\right]\hat{\psi}_{i}(x),\\
\hat{H}_{\rm I} & = & \int dx\left\{ \sum_{i=1}^{2}\left[\frac{g_{ii}}{2}\hat{\psi}_{i}^{\dagger}(x)\hat{\psi}_{i}^{\dagger}(x)\hat{\psi}_{i}\hat{\psi}_{i}(x)\right]\right.\nonumber\\
 & & \left. + g_{12}\hat{\psi}_{1}^{\dagger}(x)\hat{\psi}_{2}^{\dagger}(x)\hat{\psi}_{2}\hat{\psi}_{1}(x)\right\},
\label{eq:interactionHam}\\
\hat{H}_{\rm C} & = & 
\int dx\left\{ \hbar \frac{\delta}{2} \hat{\psi}_{2}^{\dagger}(x)\hat{\psi}_{2}(x)-\hbar \frac{\delta}{2} \hat{\psi}_{1}^{\dagger}(x)\hat{\psi}_{1}(x)\right.\nonumber\\
 & & \left.-\hbar \Omega(t) [\hat{\psi}_{1}^{\dagger}(x)\hat{\psi}_{2}(x) + \hat{\psi}_{2}^{\dagger}(x)\hat{\psi}_{1}(x)]\right\}.
\label{eq:couplingHam}
\end{eqnarray}
 The Bose field operator $\hat{\psi}_{i}(x)$ for component $i$ annihilates a particle at position $x$, and obeys the commutation relations $[\hat{\psi}_i(x),\psi^\dag_j(x')] = \delta(x-x') \delta_{ij}$.
The transversal dimensions, assumed to be harmonically trapped with frequency $\omega_{\perp}$, have been integrated out from the three-dimensional Hamiltonian, yielding the 1D interaction constants $g_{ij} = 2\hbar^2 a_{ij}/(m a_{\perp})$, where $a_{ij}$ is the scattering length and $a_{\perp} = \sqrt{\hbar/m\omega_{\perp}}$. The coupling Hamiltonian $\hat{H}_{\rm C}$ depends on the coupling strength $\Omega(t)$ and on the detuning $\delta$ between the light-field and the energy difference of the states.  Here we assume the coupling is on resonance and set $\delta=0$ for the remainder of this paper.

The nature of the ground state of a binary system is determined by the balance between the interaction constants $g_{ij}$ \cite{Ho:1996up}. If the intraspecies interactions dominate, $g_{ii}\gg g_{12}$, the energy is minimised by lowering the individual densities of both components,  which then occupy all the available volume. In this miscible phase, the two species coexist at every point.  However if the interspecies interaction dominates $g_{12}\gg g_{ii}$, the energy of the system is minimised by phase separation, which minimises the contribution of the $g_{12}n_{1}(x)n_{2}(x)$ term. We refer to this phase as the immiscible phase. By defining
\begin{equation}
\Delta = \frac{g_{11}g_{22}}{g_{12}^2},
\end{equation}
then we have \cite{Ho:1996up}
\begin{eqnarray}
\Delta > 1 & \qquad & \mbox{miscible phase},\nonumber\\
\Delta < 1 & \qquad & \mbox{immiscible phase.}\nonumber
\end{eqnarray}
However, the introduction of the coupling $\Omega$ between the states can qualitatively change this picture. In the strong coupling regime, when $\Omega$ is much larger than a threshold value $\Omega_{\rm cr}$, the ground state of the system approaches the superposition state $|\psi\rangle = (|1\rangle + |2\rangle)/\sqrt{2}$. In this situation the atoms of a naturally immiscible mixture then spatially coexist as for the miscible phase. The calculation of the value of the critical coupling $\Omega_{\rm cr}$ can be found in~\ref{app:criticalCoupling}.

\subsection{Kibble-Zurek mechanism}

To formulate the KZ mechanism~\cite{Kibble:1976vm,Zurek:1985wh} for this system, we define the dimensionless control parameter
\begin{equation}\label{eq:epsilonFlat}
\epsilon(t) = \frac{\Omega_{\rm cr} - \Omega(t)}{\Omega_{\rm cr}},
\end{equation}
which quantifies the distance of the system from the critical point $\Omega_{\rm cr}$. For a continuous phase transition in the thermodynamic limit, both the equilibrium correlation length $\xi$ and relaxation time $\tau$ of the system diverge as
\begin{eqnarray}
\xi(t) & = & \frac{\xi_{0}}{|\epsilon(t)|^{\nu}},\label{eq:corrLength}\\
\tau(t) & = & \frac{\tau_{0}}{|\epsilon(t)|^{\nu z}}\label{eq:relaxationTime}.
\end{eqnarray}
where $\nu$ and $z$ are the critical exponents of the transition and define its universality class. The constants $\xi_{0}$ and $\tau_{0}$ depend on the particular characteristics of the system, such as the atomic species, particle density, trapping frequencies, and so on. The correlation length represents the average size of the regions where the system has uniform characteristics, while the relaxation time characterises the time needed by the system to adjust to an external change. For a system \emph{dynamically} approaching a critical point,  there exists a moment when the relaxation time is equal to the characteristic time scale on which the environment is changing. The KZ mechanism predicts that beyond this time the system is not able to follow the quench and remain in quasi-equilibrium -- instead it undergoes impulsive behaviour which freezes the configuration of defects in space. We consider a quench of the coupling parameter for our system of the form
\begin{equation}
\Omega(t) = \max \left[0,2\Omega_{\rm cr}\left(1-\frac{t}{\tau_{\rm Q}}\right)\right],
\end{equation}
where $\tau_{Q}$ is the quench time.  The moment  when the relaxation time $\tau[\epsilon(\hat{t})]$ becomes larger than the characteristic time of the quench $|\epsilon(\hat{t})/\dot{\epsilon}(\hat{t})|$  determines the freezing time $\hat{t}$ through the condition
\begin{equation}\label{eq:freezingTime}
\tau[\epsilon(\hat{t})] = \epsilon(\hat{t})/\dot{\epsilon}(\hat{t}).
\end{equation}
Solving Eq.~\eqref{eq:freezingTime} with the aid of Eqs.~\eqref{eq:corrLength}--\eqref{eq:relaxationTime} for the freezing time gives
\begin{equation}
\hat{t} = \tau_{0}^{1/(1+\nu z)}\tau_{\rm Q}^{\nu z/(1+\nu z)}.
\end{equation}
The correlation length at freezing time, $\hat{\xi} = \xi_{0}(\tau_{\rm Q}/\tau_{0})^{\nu/(1+\nu z)}$, fragments the system in a series of regions with uniform characteristics. Since topological defects can form only on the boundaries between these regions \cite{Zurek:1996ik} the number of formed defects is given by
\begin{equation}\label{eq:KZscaling}
N_{\rm d} = \frac{L}{\hat{\xi}} = \frac{L}{\xi_{0}}\left(\frac{\tau_{0}}{\tau_{\rm Q}}\right)^{\frac{\nu}{1+\nu z}}.
\end{equation}

\subsection{Simulation method}

Our goal in this paper is to perform a computational study of the dynamics of the engineered quantum phase transition.  In order to do so, we must go beyond the mean-field approximation for the system.  For a system with a uniform initial density, dynamically crossing the critical coupling $\Omega_{\rm cr}$ leads to a modulational instability --- such that the initial state is dynamically unstable.  In mean-field simulations such instabilities are seeded by numerical noise.

An approximate method for simulating beyond mean-field methods for ultra-cold gases is the truncated Wigner approximation~\cite{Steel:1998vf, Sinatra:2002tx,Blakie:2008vt}.  Briefly, this is a phase space method that simulates the evolution of the Wigner function for a system by means of stochastic trajectories.  It is approximate in that the stochastic representation neglects some terms in the equation of motion involving third and higher orders derivatives of the Wigner function with respect to the phase space variables. However, these have a small contribution for short times when the number of particles in the system is much larger than the number of modes that are simulated~\cite{Sinatra:2002tx,Blakie:2008vt}.

The net result is that the equations of motion for the system are simply the appropriate Gross-Pitaevskii equations for the coupled condensates, but the initial state must be sampled from the Wigner distribution.  For a weakly interacting BEC at zero temperature, this corresponds to populating the Bogolioubov quasiparticle modes with half a particle of classical noise, which represents the initial quantum fluctuations.  Expectation values of symmetrised products of equal time quantum field operators are calculated by the averages of the corresponding phase space variable over a number of trajectories beginning with different initial conditions.

In our simulations, the initially small "quantum noise" then seeds the modulational instability, and this leads to macroscopically different outcomes as might be expected to be realised in an experiment.  Indeed, it has been plausibly argued that individual trajectories can be roughly interpreted as being equivalent to single experimental runs~\cite{Blakie:2008vt}.  This is the approach that we take for the analysis in this paper.
 
The equations of motion for the two components $\psi_i(x)$ for $i=1,2$ are
\begin{equation}\label{eq:GPE}
\fl
i\hbar\frac{\partial {\psi_{i}}(x)}{\partial t}= \left[ -\frac{\hbar^2}{2m}\frac{\partial^2}{\partial x^2}+V(x) + g_{ii}|\psi_i(x)|^2+g_{12}|{\psi}_{3-i}(x)|^2\right]{\psi}_{i}(x) - \hbar\Omega {\psi}_{3-i}(x).
\end{equation}
We construct the initial states for our simulation for the homogeneous system beginning from the even superposition $\psi_{+}(x) = [\psi_{1}(x) + \psi_{2}(x)]/\sqrt{2}$ with initial states $\psi_1(x)=\psi_2(x) = \sqrt{N/2L}$ where $N$ is the total number of particles and $L$ is the length of the ring.  For the case of an elongated BEC the initial Gross-Pitaevskii solution is found by imaginary time propagation of Eq.~\eqref{eq:GPE} in the strong coupling regime. As prescribed by the TWA, the initial state is then formed by generating the complex Gaussian noises $\eta_{i}$ with variance $\overline{\eta_{i}^{*}\eta_{j}}=\delta_{ij}/2$, and forming
\begin{equation}
\psi(x,t=0) = \psi_+(x) + \sum_i [\eta_i u_i(x) + \eta_i^* v_i^*(x)],
\end{equation}
where $[u_i(x), v_i(x)]$ are the amplitudes of the $i$-th Bogoliubov mode, found by solving the Bogoliubov-de Gennes equation for a binary BEC \cite{Tommasini:2003tx,Search:2001wz}.

\section{Homogeneous Bose-Einstein condensate in a ring trap}\label{sec:uniform}

We begin our investigation of the KZ mechanism in the engineered phase transition in a binary BEC by considering a system confined in a ring trap, as was originally described in Ref.~\cite{Sabbatini:2011gu}.  For our simulation we choose to simulate $^{87}$Rb in a ring of circumference $L = 96$ $\mu$m, and take the scattering lengths to be $a_{11}=a_{22}=a_{12}/2 = 1.325$ nm.  We note that $a_{11}$ and $a_{22}$ are close to the measured values for $^{87}$Rb, but that $a_{12}$ is approximately half the naturally occurring value, making our system strongly immiscible with $\Delta = 0.25$.  We will discuss this issue further in Sec.~\ref{sec:feasibility}.   Furthermore, we take the total number of atoms to be $N = 5\cdot 10^4$  and the transverse trapping frequency $\omega_{\perp} = 2\pi \cdot 2$ kHz which sets the spin healing length $\xi_{\rm s}=\hbar/\sqrt{2m\rho g_{\rm s}}$ to be 1.5 times the transversal size of the system --- enough to freeze the transversal spin degrees of freedom. The spin interaction constant is given by $g_{\rm s} = (g_{11}+g_{22}-2g_{12})/2$. We simulate phase transitions with quench times $\tau_{\rm Q}$ over three order of magnitudes, in the range $[0.1,125]$ ms. The large ratio between the number of particles $N$ and the number of simulated modes $M=1024$ ensures the validity of the truncated Wigner approximation \cite{Sinatra:2002tx}.

\begin{figure}
\begin{center}
\includegraphics[width=0.9\textwidth]{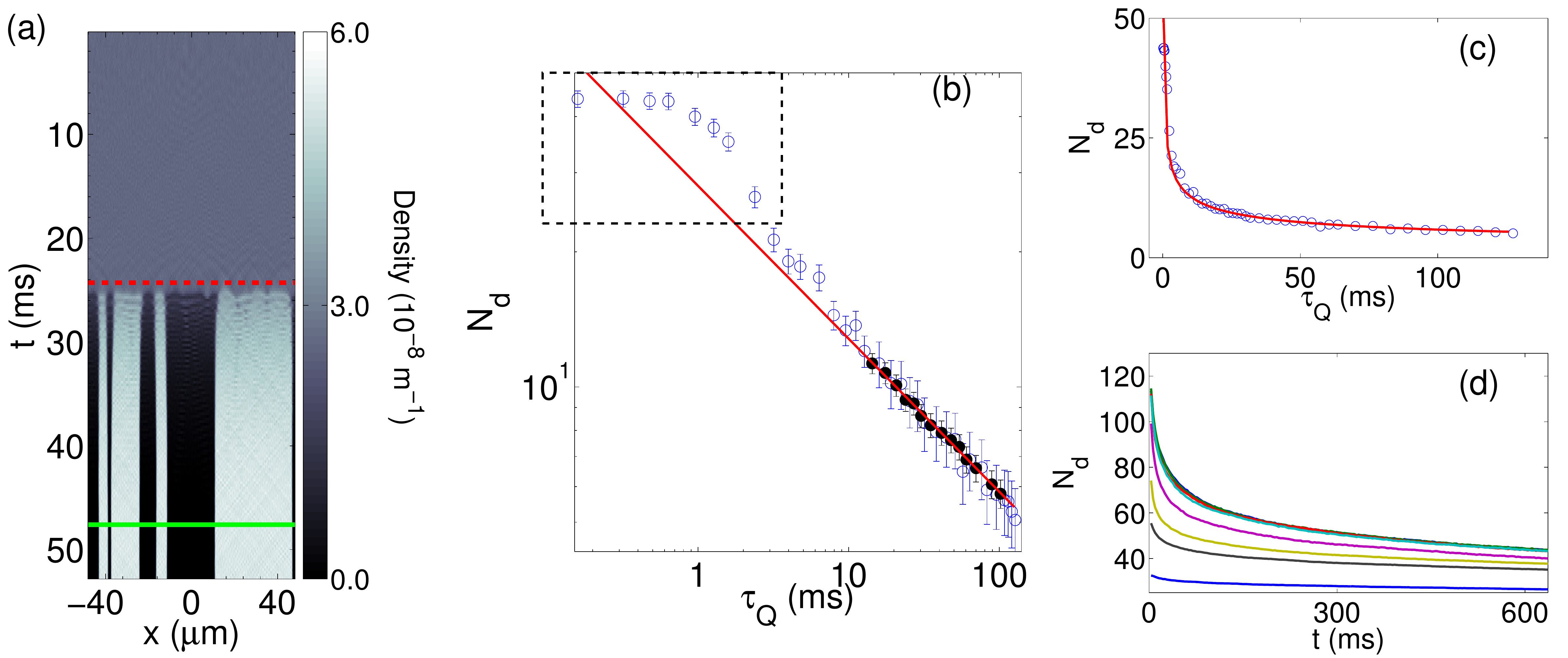}
\end{center}
\caption{Homogeneous phase transition in a ring BEC. (a) Time evolution of one component showing domain formation for $N=10^5$ and $\tau_{\rm Q}=47$ ms. The dashed (red) line represent the moment when the quench reaches the critical point $\Omega(t)=\Omega_{\rm cr}$. The solid (green) line signal the end of the quench $\Omega(t)=0$. (b) Scaling of the mean number of domains $N_{\rm d}$ in function of the quench time $\tau_{\rm Q}$ in $\log$-scale. Filled symbols represent data averaged over an ensemble of 1000 trajectories, the open symbol over 100 trajectories. Error bars derive from the standard deviation. Fitting over the filled symbols yield a scaling exponent of $n=0.341\pm 0.006$ and a factor $a_{\rm uni}=8.63\pm 0.03$. The fitted exponent $n$ shows good agreement between the numerical simulations and the KZ prediction $1/3$. (c) Same data in linear scale showing the exponential scaling. (d) Time evolution of the mean number of domains during the quench for the points in the dashed box. For fast quenches the number of domains is still decreasing at the end of our simulation as result of phase ordering.}\label{fig:uniform}
\end{figure}

The time evolution of the condensate density is shown in Fig.~\ref{fig:uniform}(a) where it is possible to observe an example of the final pattern of domains.
The number of  domains formed at the end of the simulation is identified as the number of zero crossings of the function $f(x)=|\psi_{1}(x)|^2 - |\psi_{2}(x)|^2$. This method of domain counting can easily be adopted experimentally by performing absorption imaging of the two components after Stern-Gerlach separation \cite{Hamner:2011bq,Tojo:2010iz}. The mean number of formed domains $N_{\rm d}$ versus the quench time $\tau_{\rm Q}$ is plotted in Fig.~\ref{fig:uniform}(b). We fit the power law $N_{\rm d} = a_{\rm uni}/\tau_{\rm Q}^{n}$ to the results of the simulations with $\tau_{\rm Q}>2$ ms  for which we simulated 1000 trajectories to minimise statistical uncertainty.  We find a scaling exponent of $n=0.341\pm 0.006$ while $a_{\rm uni}=8.63\pm 0.03$. The scaling exponent $n$ is in good agreement with the theoretical prediction of the KZ theory $n=1/3$, obtained from Eq.~\eqref{eq:KZscaling} with the mean-field critical exponents $\nu=1/2$ and $z=1$ as derived in~\ref{app:criticalCoupling} and in Ref.~\cite{Lee:2009ib}.

The simulation data in Fig.~\ref{fig:uniform}(b) shows a deviation from the expected scaling behaviour for fast quench times $\tau_{\rm Q}<2$ ms. As quenches become faster ($\tau_Q$ decreases), the KZ mechanism predicts a decreasing correlation length at the freezing time, and hence a smaller average domain size. In contrast, in our scheme the correlation length has a lower bound given by the spin healing length $\xi_{\rm s}$, inducing the saturation of the number of domains for fast quenches as seen in Fig.~\ref{fig:uniform}(b). The value of the saturation we extract from our data is compatible with the predicted maximum number given by $N_{\rm d}^{\rm max} \approx L/\xi_{\rm s}$.

However, we also observe a mean number of domains significantly higher than the prediction for fast quenches [dashed box in Fig.~\ref{fig:uniform}(b)]. We plot the time evolution of $N_{\rm d}$ for these quenches in Fig.~\ref{fig:uniform}(d). We note that for $\tau_{\rm Q}<1.5$ ms the number of domains is still decreasing at the end of the integration time. It is in principle possible to extend the integration time to extract the true value of $N_{\rm d}$ after stabilisation occurs and study phase ordering \cite{Biroli:2010gn} but we encountered significant numerical error and the ``thermalisation'' of the quantum noise when pushing the integration time beyond $600$ ms \cite{Sinatra:2002tx}. The latter is a known signature of the invalidity of the TWA for long evolution times, and is particularly accentuated by fast coupling quenches. These result in the rapid growth of the population of short wavelength  modes, leaving the system in a highly excited state that enhances the thermalisation process.

\section{Inhomogeneous Bose-Einstein condensate in a elongated harmonic trap}\label{sec:trappedBEC}

\begin{figure}
\begin{center}
\includegraphics[width=0.9\textwidth]{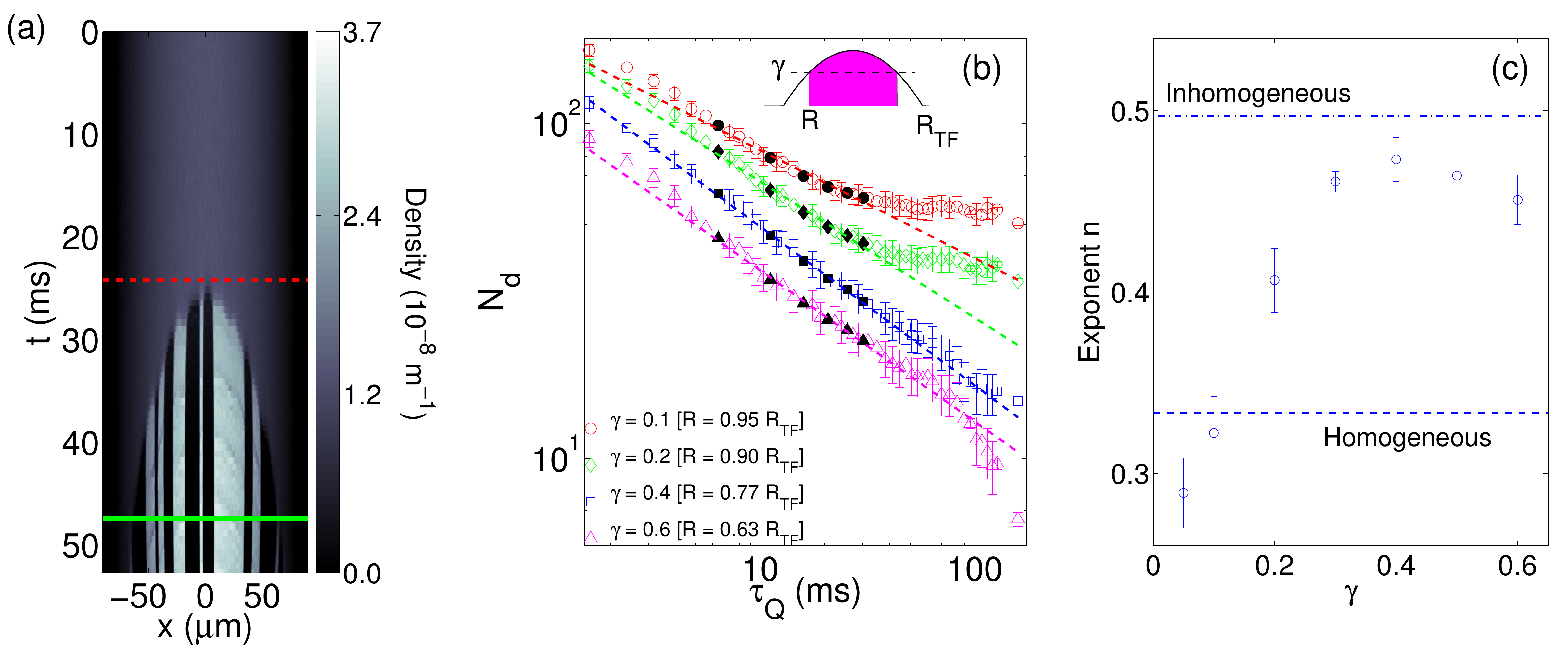}
\end{center}
\caption{(a) Time evolution of the density of one component in an harmonically trapped BEC. Dashed (red) line indicates the moment when $\Omega(t)=\Omega_{\rm cr}(0)$ (see text) while for the solid green line $\Omega(t) = 0$. The phase transition starts at the centre of the condensate where the density is highest, and propagates towards the edges. (b) Mean number of  defects formed $N_{\rm d}$ as a function of the quench time $\tau_{\rm Q}$ for several values of the counting threshold $\gamma = \rho_{\rm cut}/|\psi_{1}(0)|^2$. $R_{\rm TF}$ is the Thomas-Fermi radius at the beginning of the quench. The scaling exponents are obtained by fitting a power law to the data points for which we simulated 1000 trajectories (filled symbols). (c) Dependence of the fitting exponent on the counting region expressed by $\gamma$. The dashed line indicates the exponent for an homogeneous quench in a ring BEC. The dashed-dotted line indicates the exponent obtained by simulating an inhomogeneous phase transition in a ring geometry (see text).}\label{fig:trapped}
\end{figure}

We now move to consider the same experiment performed for a binary BEC in an elongated harmonic trap, as is common to many experimental setups for ultra-cold gases.  We take $V(x)=m\omega_{\rm x}^2 x^2 /2$ with $\omega_{\rm x}=2\pi \cdot 5$ Hz, with all other parameters the same as in Sec.~\ref{sec:uniform}.  An example trajectory for the trapped case is shown in Fig.~\ref{fig:trapped}(a).

For the homogenous BEC it was relatively straightforward to identify the location of domain walls.   For the inhomogeneous system, near the boundary of the condensate the densities $|\psi_{i}(x)|^2$ are of a comparable magnitude to the noise that is added to the initial state, and the density difference function $f(x)=|\psi_{1}(x)|^2 - |\psi_{2}(x)|^2$ can develop zeros that are not clearly domain walls.  Experimentally an equivalent issue arises due to the presence of thermal and instrumental noise in the imaging system as discussed in \cite{Weiler:2008eu} for quantum vortices.

To attempt to address this issue, we introduce a new parameter $\gamma$ that limits the counting region for domains with a particle density larger than the threshold value $\rho_{\rm cut} = \gamma \rho(0)$. Figure~\ref{fig:trapped}(b) shows the behaviour of the average number of formed domains in function of the quench time for several values of $\gamma$.
The scaling exponent is again determined by fitting a power law to the data points for quench times where we simulated 1000 trajectories. Figure~\ref{fig:trapped}(c) plots the extracted scaling exponent $n$ as a function of $\gamma$. Here we can see that $n$ shows a dependence on the counting region with values ranging from $n=0.289\pm 0.019$ to $n=0.473\pm 0.012$, but is a relatively constant  $n\approx 0.47$ for $\gamma>0.3$, pointing to a systematic difference of the scaling exponent for an harmonically trapped BEC with respect to the homogeneous case. For large counting regions $\gamma>0.25$ (where domains are counted in the low density region of the condensate), it can be seen in Fig.~\ref{fig:trapped}(b) the number of domains  for larger quench times $\tau_{Q}$ seems to converge to a value larger than predicted by a scaling law. This effect can be traced to the miscounting of noise as domains which results in an overestimate of the physical number of domains $N_{\rm d}$.

The emergence of a different scaling exponent for $\gamma > 0.3$ in the harmonically trapped case can be linked to the effects of inhomogeneity and causality, and this is the major topic we wish to address in this paper. Qualitatively, the value of the critical coupling is connected to the particle density, therefore the phase transition in an harmonically trapped BEC becomes inhomogeneous. The centre of the BEC, the point of highest density, enters the new phase first and the transition proceeds then towards the edges, as can be seen in  Fig.~\ref{fig:trapped}(a) where the domain formation near the edges occurs at later times, and a propagating phase front can clearly be identified.  In a trapped system the spatial dependence of $\Omega_{\rm cr}(x)$ and consequently of the control parameter $\epsilon_{\rm tr}(x,t)$ breaks the translational symmetry giving a preferred direction of motion for the newly formed domains. In systems of finite size this has the effect of increasing the annihilation rate of domains or the rate at which they escape our counting regions. This effect becomes increasingly important for slower quenches, and increases the observed scaling exponent.

\subsection{Causality in the harmonically trapped system}

\begin{figure}
\begin{center}
\includegraphics[width=0.8\textwidth]{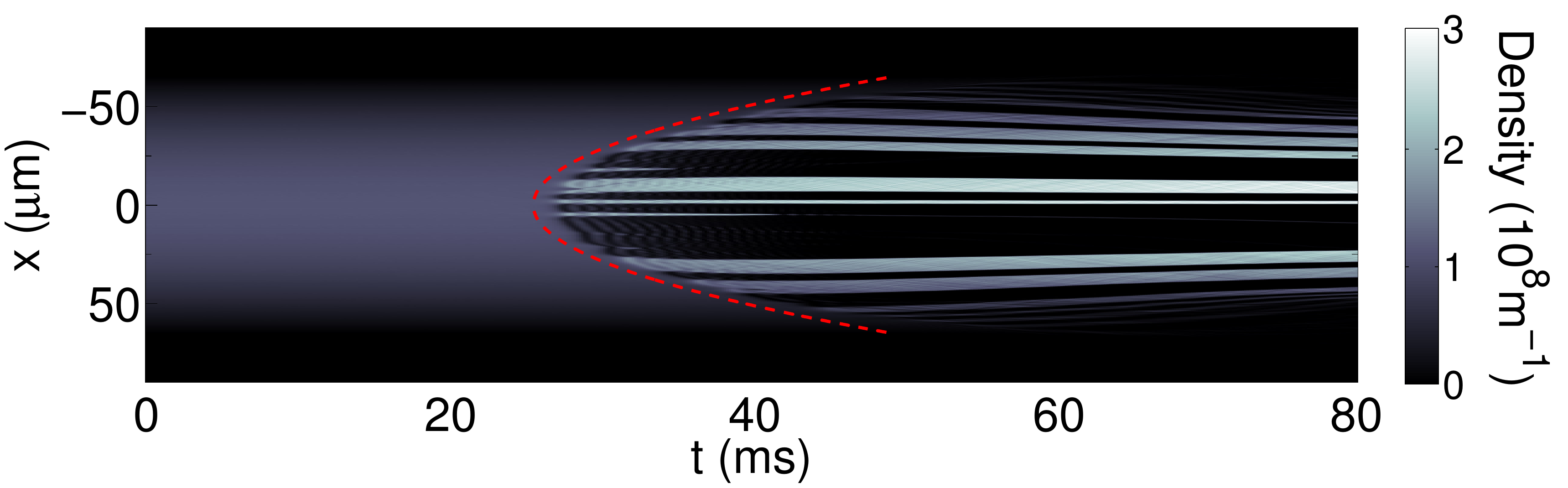}
\end{center}
\caption{Evolution of the density of one component for a trajectory with $\tau_{\rm Q} = 50$ ms. The red dashed line indicates the propagating front of the phase transition predicted by Eq.~\eqref{eq:front}.}\label{fig:front}
\end{figure}

The issue of causality is  related to the speed of the front separating regions in the unbroken (old) and broken (new)  symmetry phase. It has been proposed \cite{Zurek:2009wx,Kibble:1997gt,Dziarmaga:1999vw} that in the case where the front is \emph{slower} than the speed of sound, information can propagate from regions in the new phase to regions in the old phase. This exchange of information influences the choice of symmetry of the system in the old phase, reducing the number of defects that are eventually formed. To analyze this phenomenon, we start by deriving an expression for the density of a binary condensate in the strong coupling regime. Applying the Thomas-Fermi (TF) approximation for the density~\cite{pethicksmith} we have
\begin{equation}\label{eq:rhoTF}
\rho_{1}(x)=\rho_{2}(x)\equiv \rho(x)=\frac{\mu -V(x) + \hbar\Omega(0)}{g_{12}+g},
\end{equation}
for $x$ smaller than the Thomas-Fermi radius $R_{\rm TF}$ defined by the solution of $\mu -V(R_{\rm TF})+\hbar\Omega(0)=0$, where $\mu$ is the chemical potential resulting from $N =2\int dx\rho(x)$. As described above, the critical coupling $\Omega_{\rm cr}$ is now spatially inhomogeneous, with dependence given by $\hbar\Omega_{\rm cr}(x) = (g_{12}-g)\rho(x)$. We can modify our definition of the control parameter Eq.~\eqref{eq:epsilonFlat} to reflect the spatial inhomogeneity of the phase transition
\begin{equation}\label{eq:trappedEpsilon}
\epsilon_{\rm tr}(x,t) = \frac{\Omega_{\rm cr}(x) - \Omega(t)}{\Omega_{\rm cr}(x)}= 1 - 2\frac{\rho(0)}{\rho(x)}\left(1-\frac{t}{\tau_{\rm Q}}\right),
\end{equation}
where  we have used $\Omega(t) = \max[0,2\Omega_{\rm cr}(0)(1-t/\tau_{\rm Q})]$. In inhomogeneous phase transitions, different regions of the system experience different effective quench time as a consequence of the spatially dependent critical coupling. The effective quench time is again defined as the change rate of the newly defined control parameter $\epsilon_{\rm tr}(x,t)$,
\begin{equation}\label{eq:tauQx}
\tau_{\rm Q}(x) = \left|\frac{\partial \epsilon_{\rm tr}(x,t)}{\partial t}\right|^{-1} = \tau_{\rm Q}\frac{\mu - V(x) + \hbar\Omega(0)}{2[\mu + \hbar\Omega(0)]}.
\end{equation}
We note that the effective quench time $\tau_{\rm Q}(x)$ approaches zero for $x\to R_{\rm TF}$. Therefore, outer regions of the BEC are subject to a smaller correlation length at freezing time, and the average size of the domains decrease in the proximity of the condensate edges, see Figs.~\ref{fig:trapped}(a) and \ref{fig:front}.

Solving the equality $\epsilon_{\rm tr}(x_{\rm F},t_{\rm F})=0$ for $x_{\rm F}$ defines the trajectory of the front which in the case of harmonic potential moves according to
\begin{equation}\label{eq:front}
x_{\rm F}(t) = \sqrt{ \frac{2}{m\omega^2} [\mu + \hbar\Omega(0)] \left( \frac{2t}{\tau_{\rm Q}}-1\right) }.
\end{equation}
We have verified that Eq.~\eqref{eq:front} accurately reproduces the trajectory of the front by comparing it with the simulations, as shown in Fig.~\ref{fig:front}. The speed of the front during the transition follows from Eq.~\eqref{eq:front} as
\begin{equation}
v_{\rm F}(x_{\rm F}) = \left| \frac{dx_{\rm F}}{dt} \right| = \frac{2[\mu + \hbar \Omega(0)]}{m\omega_{\rm x}^2 \tau_{\rm Q}x_{\rm F}}.
\end{equation}

\subsection{Kibble-Zurek scaling exponent for the harmonically trapped system}

\begin{figure}
\begin{center}
\includegraphics[width=0.9\textwidth]{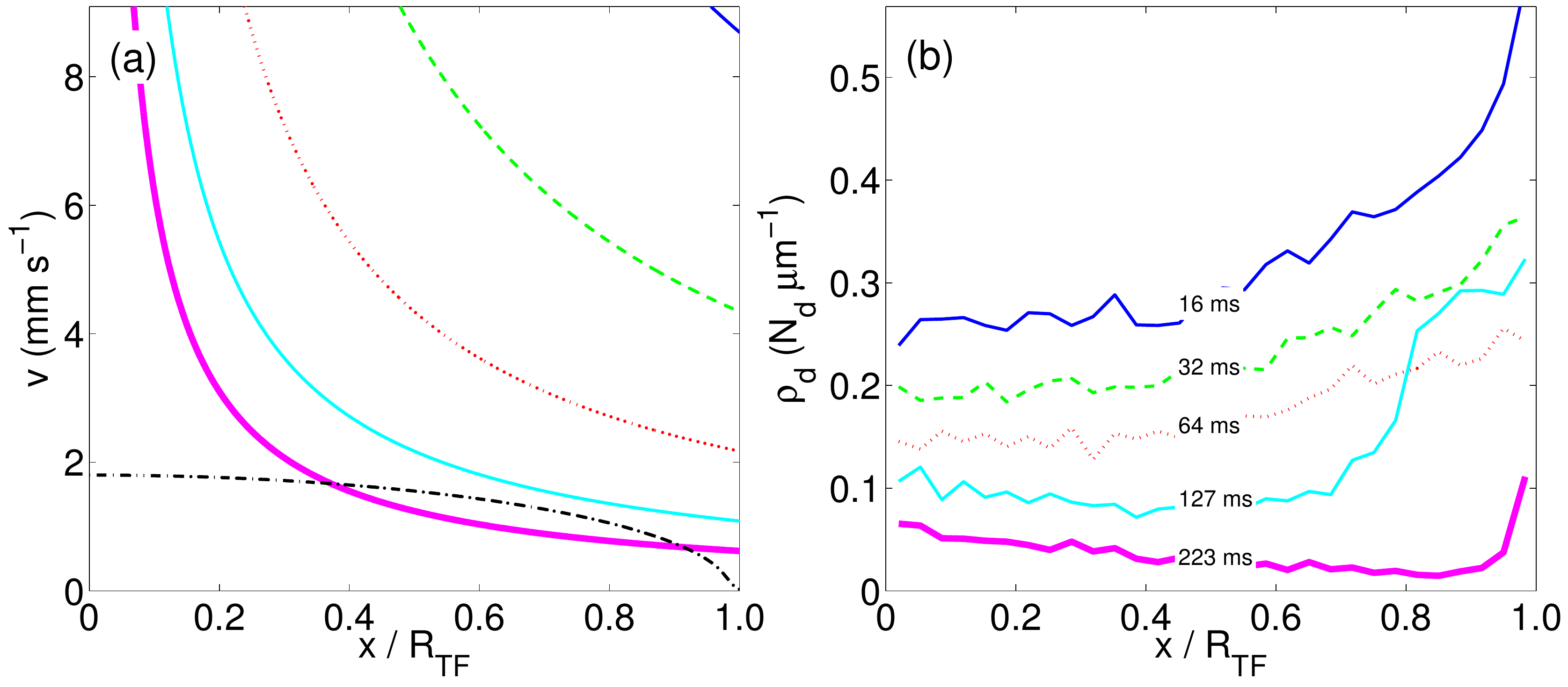}
\end{center}
\caption{(a) Comparison between the speed of sound (dot-dashed line) and the speed of front propagation for the quench times expressed in (b). (b) Final spatial density of defects for the trapped system for a range of quench times $\tau_{\rm Q}$ as indicated in the figure. $R_{\rm TF}$ is the Thomas-Fermi radius of the condensate.}
\label{fig:spatialDomains}
\end{figure}

In order for the new phase to transfer information to the old phase, the front speed has to be smaller than the relevant speed of sound at freezing time given by $\hat{v}=\hat{\xi}/\hat{\tau}=\xi_{0}/[\tau_{0}^{(1+\nu)}\tau_{\rm Q}^{\nu(z-1)}]^{\frac{1}{1+\nu z}}$ \cite{Zurek:2009wx} where we have used $\hat{\xi}=\xi_{0}\epsilon(\hat{t})^{-\nu}=\xi_{0}[\tau_{\rm Q}/\tau_{0}]^{\frac{\nu}{1+\nu z}}$ and $\hat{\tau}=\tau_{0}\epsilon(\hat{t})^{-\nu z}=[\tau_{0}\tau_{\rm Q}^{\nu z}]^{\frac{1}{1+\nu z}}$. For the phase transition under consideration here ($z=1$), $\hat{v}$ is independent of the quenching time and equal to the local speed of sound $v_{\rm s}(x) = \sqrt{(g+g_{12})\rho(x)/(2m)}$ \cite{Jenkins:2003jk,Tommasini:2003tx}. Solving the inequality
\begin{equation}\label{eq:causalityBoundary}
\frac{2[\mu + \hbar \Omega(0)]}{m\omega_{\rm x}^2 \tau_{\rm Q}x}\le \sqrt{\frac{(g+g_{12})\rho(x)}{2m}},
\end{equation}
for $x$ defines the region $\hat{X}$ within which the formation of defects is suppressed. For the parameters used in this work Eq.~\eqref{eq:causalityBoundary} has no solution for $\tau_{\rm Q}<\tau_{\rm Q}^{\rm cr}\approx 150$ ms. Thus for a large range of quench times we simulate, the front is always faster than the sound, and the phase transition is not affected by the broken symmetry choices in the neighbouring regions. For $\tau_{\rm Q}>\tau_{\rm Q}^{\rm cr}$ on the other hand, the speed of the front equals the speed of sound in two points (see Fig.~\ref{fig:spatialDomains}a, where  the pink solid line compares the speed of the phase transition front to the local speed of sounds for $\tau_Q$ = 223 ms.)  When the front first enters the condensate at the centre  it moves faster than the sound for a short distance, until it slows into a subsonic regime. Approaching the edges, where the speed of sound goes to zero, the front again becomes supersonic. As the size of the region with suppressed domain formation increases with the increasing quench time, the effect of causality on inhomogeneous phase transitions is to \emph{increase} the scaling exponent, by further reducing the total number of domains formed compared to the fully supersonic case. We can compute the number of formed domains in this scenario by integrating the effective quench time $\tau_{\rm Q}(x)$ over the supersonic region:
\begin{equation}\label{eq:Ncausality}
N_{\rm d} \propto \int_{0}^{\hat{x}_{1}} dx \tau_{\rm Q}(x)^{-\nu/(1+\nu z)} + \int_{\hat{x}_{2}}^{R_{\rm TF}} dx \tau_{\rm Q}(x)^{-\nu/(1+\nu z)},
\end{equation}
where $\hat{x}_{1}$ and $\hat{x}_{2}$ are respectively the  inner and outer boundaries of the subsonic region $\hat{X}$. When integrated numerically with $\nu = 1/2$ and $z=1$, Eq.~\eqref{eq:Ncausality} yields a scaling exponent $n\approx 0.37$. 

In more detail, we note that the comparison of the local speed of sound to the front propagation speed does not provide a rigorous criterion for delimiting the suppression of domain formation. In fact, the authors of Ref.~\cite{Kibble:1997gt,Dziarmaga:1999vw} note that defect formation can occur for $v_{\rm F}/v_{\rm s} \geq 0.5 $. However, the introduction of any proportionality constant in Eq.~\eqref{eq:causalityBoundary} modifies the value of the critical quench time $\tau_{\rm Q}^{\rm cr}$, whereas the value of the scaling exponent remains $n \approx 0.37$. We address the issue of the critical value for the speed of the front in Sec.~\ref{sec:causality}.

\subsection{Spatial density of defects}

We turn now to the analysis of the dependence of the spatial density of defects on the condensate density for a harmonically trapped BEC. A comparison between the front speed for a range of quench times and the local speed of sound is shown in Fig.~\ref{fig:spatialDomains}(a). In Fig.~\ref{fig:spatialDomains}(b) we plot the mean spatial distribution of the defect density at the end of the integration time for the same quench times as in Fig.~\ref{fig:spatialDomains}(a). We can see that for fast quenches (for which the front speed is always larger than the speed of sound) the domain density increases towards the edge of the condensate as suggested by Eq.~\eqref{eq:tauQx}. The rise in defect density for $x > 0.8 R_{\rm TF}$ is due to the difficulty of distinguishing noise from domains in the low density region, as described in Sec.~\ref{sec:trappedBEC}. Figure \ref{fig:spatialDomains}(b) shows that for slower quenches, the density of domains begins to \emph{decrease} for increasing $x$, rather than increasing as is observed for faster quenches. This is likely due to the suppression of domain formation as described above. The onset of the suppression is expected to result in a sudden and sharp decrease in the number of domains. However, we have been unable to observe this clearly in the simulation results, as the positions of the domains are not fixed when $\Omega(t) > 0$, and so domains initially seeded in the supersonic region are able to move into the subsonic region before they become clearly distinguishable.

\section{Inhomogeneous phase transition simulated in the ring geomety}\label{sec:iptRing}

While the simulations of the experiment in the harmonic  trap suggest a scaling exponent for the number of defects of $n\sim 1/2$, this is still open to question.  To confirm the modified scaling exponent derived in Sec.~\ref{sec:trappedBEC} we now "simulate" the inhomogeneous phase transition of the harmonically trapped BEC. We acheive this by performing numerical simulations of a binary BEC in a ring trap subject to a spatially dependent quench of the coupling strength designed to reproduce the physics of the trapped system. We use
\begin{equation}
\Omega(x,t) = \max\left[0,2\Omega_{cr}\frac{\rho(0)}{\rho(x)}\left(1-\frac{t}{\tau_{\rm Q}}\right)\right],
\end{equation}
where  $\rho(x)$ is the Thomas-Fermi density of the trapped BEC system we are ``simulating'' [Eq.~\eqref{eq:rhoTF}], and the circumference of the ring $L = 110$~$\mu$m~$<2R_{\rm TF}$ (equivalent to $\gamma = 0.15$) so that we avoid any divergence of $\Omega(x,t)$ at the edge of the simulated system where $\rho(x) \rightarrow 0$. In this situation the propagating phase transition front is due to the spatially dependent coupling parameter  $\Omega(x,t)$, rather than the spatially dependent density.  By design, this simulation has the same  control parameter  for the quench as $\epsilon_{\rm tr}(x,t)$ in Eq.~\eqref{eq:trappedEpsilon}.  However, we avoid the domain-counting problem described in Sec.~\ref{sec:trappedBEC} as the BEC has constant density around the ring.

\begin{figure}
\begin{center}
\includegraphics[width=0.8\textwidth]{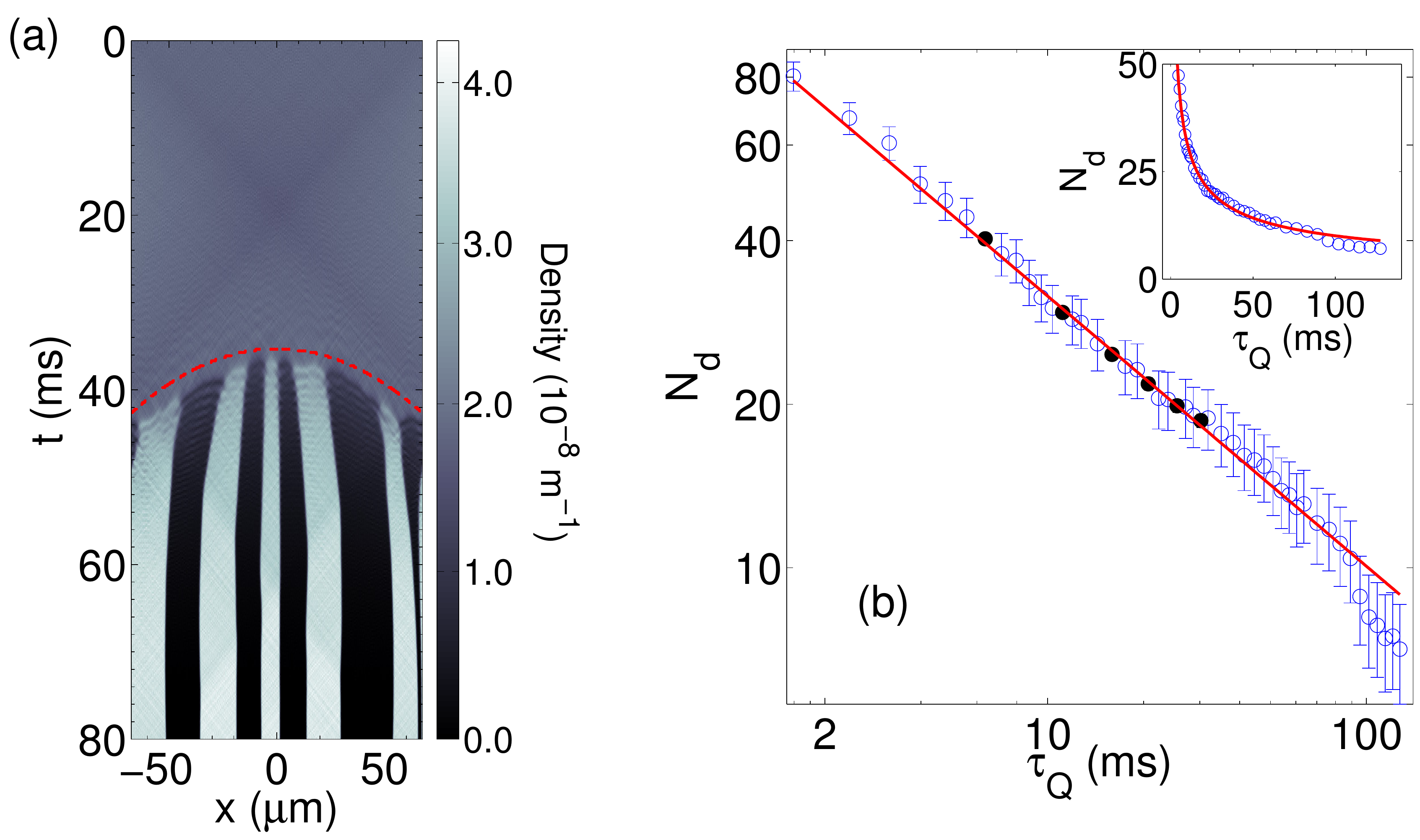}
\end{center}
\caption{(a) Evolution of the density of one component during an inhomogeneous coupling quench in a ring trap with $\tau_{\rm Q}=75$ ms. The red dashed line shows the approximate motion of the front of the phase transition from the solution of $\epsilon_{\rm ir}(x,t)=0$. (b) Scaling of the mean number of domains $N_{\rm d}$ as a function of the quench time $\tau_{\rm Q}$ on a log-scale. The fitting with a power law results in an exponent $n=0.497\pm 0.015$. The inset shows the same data on a linear scale.}
\label{fig:ringInhomogeneous}
\end{figure}

The evolution of the density of one component of the BEC is shown in Fig.~\ref{fig:ringInhomogeneous}(a). The critical coupling is first reached at $x=0$, and the front of the phase transition moves around the ring in exactly the same manner as in the trapped BEC with a spatially constant coupling. The scaling of defect number for this simulation is shown in Fig.~\ref{fig:ringInhomogeneous}(b), and we find a scaling exponent of $n=0.497\pm0.015$.  Thus we conclude that the exponent of $n\sim 0.5$ that was determined in the harmonic trap is robust.

\subsection{Kibble-Zurek scaling exponent for the inhomogeneous phase transition}
\label{sect:ringexponent}
We now estimate the Kibble-Zurek scaling exponent expected for the inhomogeneous phase transition in a ring BEC. For a ring BEC the condition Eq.~\eqref{eq:causalityBoundary} for domain suppression due to causality becomes 
\begin{equation}\label{eq:suppressionlimit}
\frac{2[\mu+\hbar\Omega(0)]}{m\omega^2\tau_{\rm Q}x_{\rm F}}\le\sqrt{\frac{(g+g_{12})N}{2mL}},
\end{equation}
where on the right hand side we have inserted the constant speed of sound in the ring \cite{Tommasini:2003tx}. The number of domains scales as function of the quench time as
\begin{equation}\label{eq:Nxuniform}
N_{\rm d}\propto \int_{-\hat{x}}^{\hat{x}}dx\,\tau_{\rm Q}(x)^{-\nu/(1+\nu z)},
\end{equation}
where $\hat{x}$ is the solution to the equality of Eq.~\eqref{eq:suppressionlimit}. If the size of the system $L$ is such that the inequality Eq.~\eqref{eq:suppressionlimit} is never satisfied, the temporal dependence in Eq.~\eqref{eq:Nxuniform} factors out.  This will increase the number of domains compared to the case of the homogeneous transition, but leaves the scaling exponent unchanged at $n=1/3$.  However, if there is a region where the speed of the phase transition front is less than the speed of sound, the defect-suppression region grows with the quench time and the scaling exponent is $n=4/3$.

For the results we present in Fig.~\ref{fig:ringInhomogeneous}(b) the quench times are small enough that there is no region of domain suppression.  However, we find the scaling exponent to be $n\approx 0.5$,  in disagreement with the prediction above.  We believe  this discrepancy is caused by the breaking of translational invariance, as we describe below in Sec.~\ref{sect:trans_inv}.

\subsection{Translational invariance and domain annihilation}
\label{sect:trans_inv}
For a BEC in a ring trap the system is symmetric under rotations about the center of the ring. This is reflected in the fact that the Hamiltonian Eq.~\eqref{eq:fullHam} of the main text is invariant under translations $x\to x'+\alpha$ when $V(x)=0$ (the definition of a homogeneous system). However, harmonically trapped condensates and spatially inhomogeneous quenches of the coupling break this symmetry.  To clearly demonstrate the effect of breaking translational invariance on the phase transition we simulate both a homogeneous and an inhomogeneous quench in a uniform system, but seed the domain formation ``by hand.''

\begin{figure}
\begin{center}
\includegraphics[height=0.15\textheight]{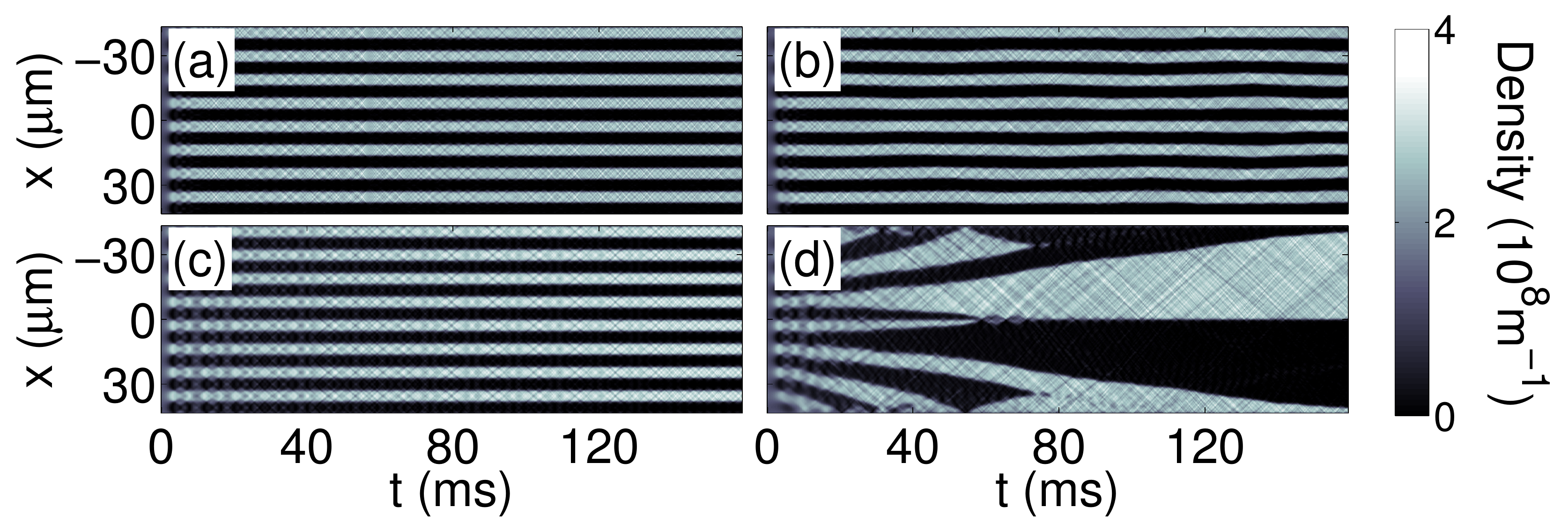}\includegraphics[height=0.15\textheight]{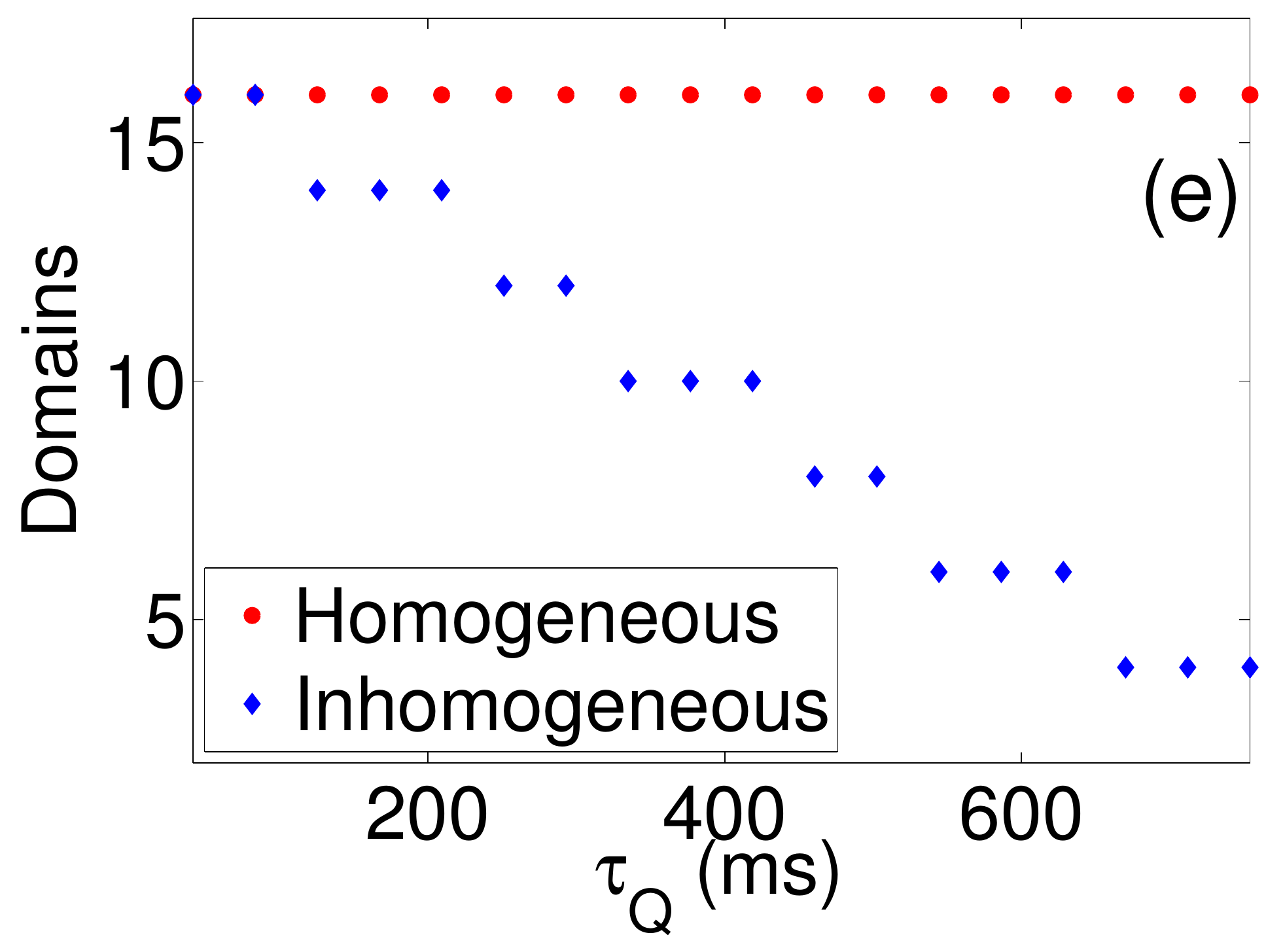}
\caption{Effect of inhomogeneity on the decay of domains. (Left panels a--d) Evolution of the condensate density for one component in a ring trap with 16 seeded domains.  Left column (a,c) is for the the homogeneous quench given by Eq.~\eqref{eq:hq}, and the right column (b,d) is for the inhomogeneous quench given by Eq.~\eqref{eq:iq}. The top row (a,b) is for a fast quench $\tau_{\rm Q}=15$ ms, where it can be seen that there is no change to the domain pattern that is seeded.  The bottom row (c,d) is for a slow quench $\tau_{\rm Q}=286$ ms.  It can be seen in (d) that there is significant dynamics and domain annihilation for the inhomogeneous quench. (e) The number of domains surviving an inhomogeneous (blue curve) and a homogeneous (red curve) quench in function of the quench time $\tau_{\rm Q}$.}
\label{fig:TranslationInvariance}
\end{center}
\end{figure}

We begin with the initial state
\begin{equation}\label{eq:seeded}
[\psi_{1}(x),\psi_{2}(x)] = \sqrt{\frac{N}{2L}}\left[1+A\sin \left(k\frac{2 \pi x}{L}\right),1-A\sin \left(k\frac{ 2 \pi x}{L}\right)\right],
\end{equation}
where $k$ is an integer that determines the number of seeded domains, and $A=0.1$ is the amplitude of the seed.  The coupling is
\begin{equation}\label{eq:hq}
\Omega_{\rm H}(x,t) = \max\left[0,2\Omega_{\rm cr} \left(0.3-\frac{t}{\tau_{\rm Q}}\right)\right],
\end{equation} 
for the homogeneous quench and 
\begin{equation}\label{eq:iq}
\Omega_{\rm I}(x,t) = \max\left[0,2\Omega_{\rm cr}\frac{\rho(0)}{\rho(x)} \left(0.3-\frac{t}{\tau_{\rm Q}}\right)\right],
\end{equation} 
for the inhomogeneous quench emulating the phase transition in the harmonically trapped BEC. Both choices ensure that $\Omega(x,0)<\Omega_{\rm cr}$ everywhere at $t=0$. The domains will grow from the seed with no propagating front for the  phase-transition, but the nonzero coupling will  affect the domain dynamics. 

We plot typical results in Fig.~\ref{fig:TranslationInvariance}. For fast quenches we find that the domains do not have time to move or decay, and the number of  domains observed at the end of the integration time is the same as the number seeded for both homogeneous and inhomogeneous quenches. However, for slow quenches we find that  $\Omega_{\rm I}(x,t)$ results in a systematically smaller number of domains than $\Omega_{\rm H}(x,t)$ for the same quench time, due to the motion and subsequent annihilation of domain walls.  This effect will increase the scaling exponent for the number of domains, and for our parameters it increases it to $n\approx 0.5$.

\section{Simulating a Homogeneous Phase Transition in an Harmonically Trapped BEC}\label{sec:rectifying}

The success of the quantum ``simulation'' in Sec.~\ref{sec:iptRing} suggests that we can potentially engineer a spatially-dependent coupling $\Omega(x,t)$ such that the phase transition in an harmonically trapped BEC happens everywhere simultaneously --- that is, the phase transition is \emph{homogeneous}. We describe our efforts to realise this below.

A harmonically trapped BEC reacts to a coupling quench by changing its shape. This means that a coupling quench of the form $\Omega(x,t) = \max[0,2\Omega_{\rm cr}^{0}(x)(1-t/\tau_{\rm Q})]$, where $\Omega_{\rm cr}^{0}(x)$ is the initial critical coupling, will not produce an homogeneous quench as it fails to take into account the change of shape of the system density. We extend Eq.~\eqref{eq:rhoTF} to take into account a spatially and time dependent coupling with the following ansatz for the density
\begin{equation}\label{eq:rhoHomTrapped}
\rho(x,t) = \frac{\mu(t)-V(x)+\hbar\Omega(x,t)}{g_{12}+g}.
\end{equation}
This reflects Eq.~\eqref{eq:rhoTF} in the case where the coupling is spatially inhomogeneous and uses the local density approximation, i.e. the condensate density is influenced only by the \emph{local} value of the trapping potential and coupling between the components. We also assume the coupling $\Omega(x,t)$ to be proportional to the critical coupling at any time
\begin{equation}\label{eq:propCoupling}
\hbar\Omega(x,t) = c(t)\hbar\Omega_{\rm cr}(x) = c(t)[g_{12}-g]\rho(x,t),
\end{equation}
where we have used the dependence of $\Omega_{\rm cr}(x)$ on the density. We have also introduced the function $c(t) = 2(1-t/\tau_{\rm Q})$ which describes the proportionality between $\Omega(x,t)$ and $\Omega_{\rm cr}(x)$. Eqs.~(\ref{eq:rhoHomTrapped}--\ref{eq:propCoupling}) constitute a system of equations describing the mutual relation between the condensate density $\rho(x,t)$ and the coupling $\Omega(x,t)$ that we set to a time-varying multiple of the critical coupling $\Omega_{\rm cr}(x,t)$ decided by $\rho(x,t)$. The condition expressed by Eq.~\eqref{eq:propCoupling} ensures the simultaneous crossing of the critical point for $c(t)=1$.

\begin{figure}
\begin{center}
\includegraphics[width=0.8\textwidth]{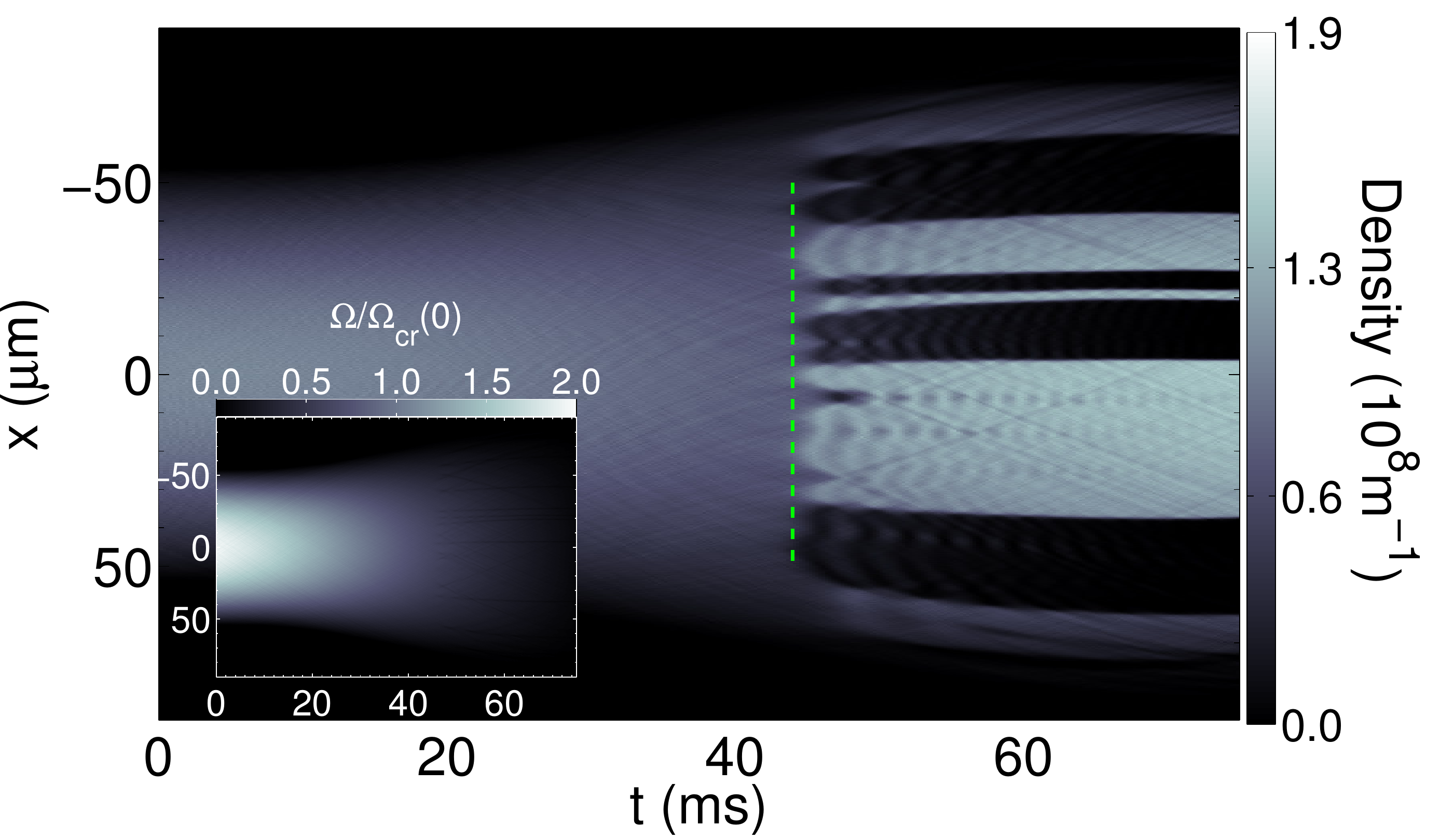}
\caption{Simulation of a homogeneous phase transition in a trapped BEC. Time evolution of the particle density of one component for a quench time of $\tau_{\rm Q}\approx 79$ ms and $N=20000$. The dashed green line denotes the region where the phase transition is homogeneous. The inset shows the time evolution of the coupling $\Omega(x,t)$.}
\label{fig:HomTrapped}
\end{center}
\end{figure}

Substituting Eq.~\eqref{eq:propCoupling} into Eq.~\eqref{eq:rhoHomTrapped} and solving the equation $\int dx \rho(x)=N/2$ we obtain an equation for $\Omega(x,t)$ for an harmonically trapped system
\begin{equation}\label{eq:omegaHomTrapped}
\hbar\Omega(x,t) = \frac{c(t)(g_{12}-g)}{G(t)}
\left[\left(\frac{3N}{8}\sqrt{\frac{m\omega_{\rm x}^2}{2}}G(t)\right)^{2/3}-\frac{m\omega_{\rm x}^2 x^2}{2}\right],
\end{equation}
where $G(t)=g_{12}+g-c(t)(g_{12}-g)$. We have implemented the coupling quench of Eq.~\eqref{eq:omegaHomTrapped}, and the evolution of the density of one of the components is  shown in Fig.~\ref{fig:HomTrapped}. We find that the phase transition is effectively homogeneous in the range $|x| < 50$~$\mu$m. However, in deriving Eq.~\eqref{eq:omegaHomTrapped} we have assumed that the condensate adiabatically follows the spatial dynamics of the quench at any time, i.e. the quench time is much larger than the condensate reaction time. For faster quenches the condensate fails to follow the changes in the coupling and the conditions leading to Eq.~\eqref{eq:omegaHomTrapped} break down. As $\tau_{\rm Q}$ decreases, the region where the transition is effectively homogeneous rapidly shrinks or even disappears, complicating the process of  domain counting. This leaves us with a narrow range of quench times $\tau_{\rm Q}$ with a homogeneous transition, which is insufficient to reliably extract a scaling exponent.

From Fig.~\ref{fig:HomTrapped}, it is also evident that the quench of Eq.~\eqref{eq:omegaHomTrapped} results in the onset of breathing dynamics that further complicates the counting process. In order to explain the origin of the breathing mode we derive the Thomas-Fermi radius of the condensate in the approximation given by Eqs.~(\ref{eq:rhoHomTrapped}-\ref{eq:propCoupling}). Inserting the solution of the system of equations into the normalisation condition $N=2\int_{\rm -R_{\rm TF}}^{\rm R_{\rm TF}}\rho(x)dx$, we obtain the chemical potential
\begin{equation}
\mu = \left[\frac{3}{8} g_{12} \left( 1 - c(t) + \sqrt{\Delta} - c(t)\sqrt{\Delta}\right) \right]^{2/3} \left(\frac{m\omega^2}{2}\right)^{1/3} N^{2/3}
\end{equation}
and the Thomas-Fermi radius
\begin{equation}\label{eq:Rtf}
R_{\rm TF} = \left[\frac{3}{8} g_{12} \left( 1 - c(t) + \sqrt{\Delta} - c(t)\sqrt{\Delta}\right) \right]^{1/3} \left(\frac{m\omega^2}{2}\right)^{-1/3} N^{1/3}.
\end{equation}
From \eqref{eq:Rtf} it can be seen that $R_{\rm TF}$ increases as $c(t)\to 1^{+}$, explaining the breathing in the case of a coupling quench from $c(0)>1$ to $c(t_{\rm final})=0$.

\section{Threshold Value for Causality}\label{sec:causality}

Previous analysis of the effect of causality on the formation of defects found that defect suppression occurs when the speed of the front is a \textit{fraction} of the speed of sound \cite{Kibble:1997gt,Dziarmaga:1999vw}. In this section we determine the value of the critical velocity of the phase front below which the formation of domains is suppressed for the coupled binary BEC. To do so we simulate the inhomogeneous phase transition in a ring with a front moving with the velocity $v_{\rm F}$. We control the velocity of the front by using the following quench of the coupling
\begin{equation}
\fl \hbar\Omega(x,t) = \max\left[0,\Omega_{\rm cr} \left\{ 2 + \tanh \left( x - \frac{w}{2} - v_{\rm F}t\right) + \tanh \left(-x-\frac{w}{2} - v_{\rm F}t\right) \right\} \right],
\end{equation}
where the initial distance between the fronts $w$ is introduced to avoid the effect of the front entering the system. We simulate this system using the same parameters as Sec.~\ref{sec:uniform} with $w=L/20$ for a time of $t = 150$ ms.

\begin{figure}
\begin{center}
\includegraphics[width=0.8\textwidth]{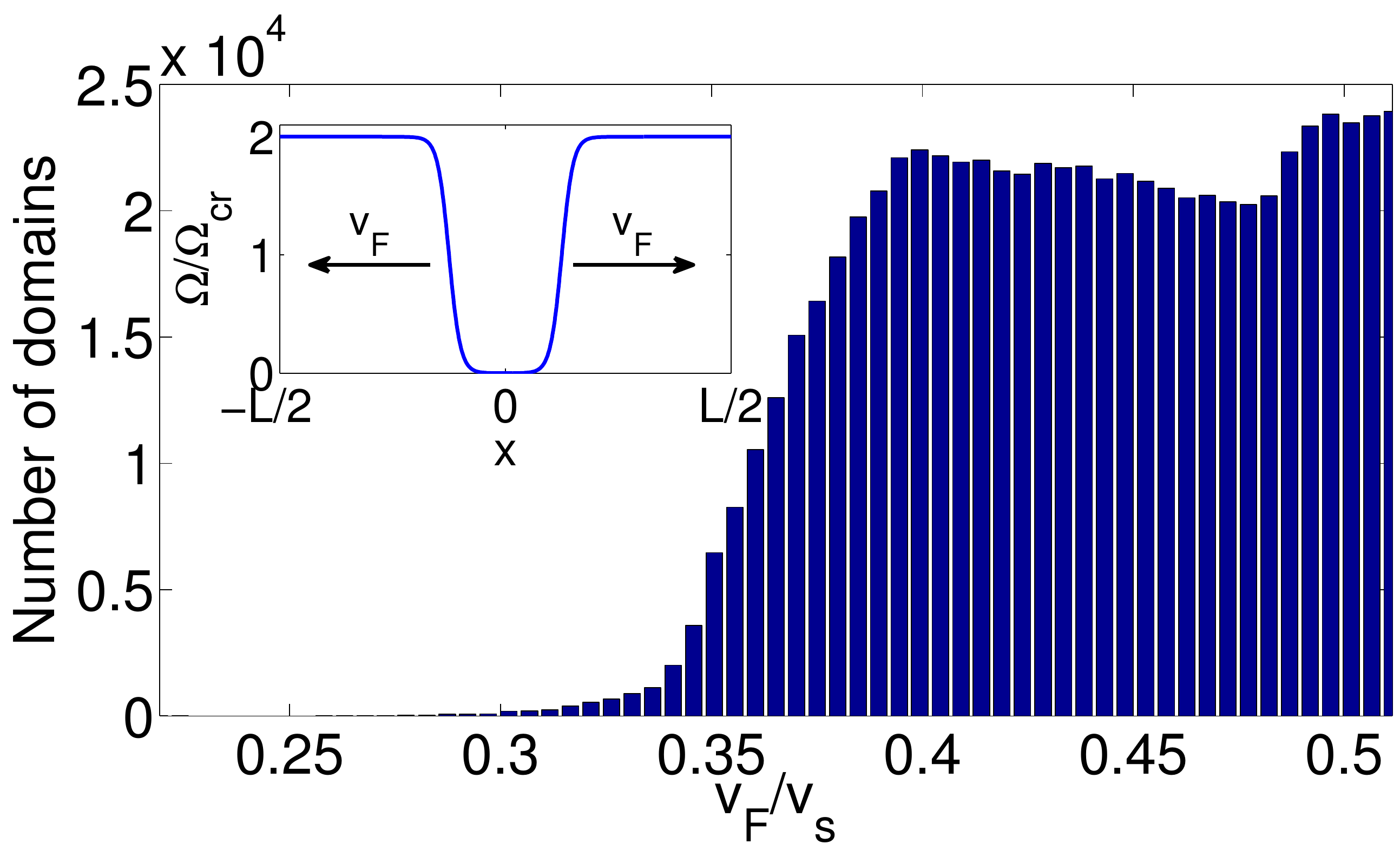}
\caption{Determination of the critical speed for domain suppression.  The mean number of  domains  formed is plotted as a function of the speed of the phase transition front $v_{\rm F}$, resulting in a critical speed of $v_{\rm F}/v_{\rm s}\approx 0.37$. The inset shows the coupling quench used to test the critical velocity.}
\label{fig:Causality}
\end{center}
\end{figure}

The mean number of defects formed  as a function of the speed of the front $v_{\rm F}$ is shown in Fig.~\ref{fig:Causality},  which indicates a value for the critical speed of $v_{\rm F}/v_{\rm s}\approx 0.37$. In the mean field approximation the transition between the supersonic and subsonic regime is expected to occur at a precise fraction of the speed of sound. However the presence of fluctuations in the density of the system means that there is some uncertainty in the the  local speed of sound, smearing out the exact value of the transition. In Ref.~\cite{Dziarmaga:1999vw} the authors determine the relevant speed for the onset of causality using a moving step function for the control parameter. In this work we use a spatially continuous control parameter and consequently the phase transition occurs on a finite width. This finite width is however small and comparable with the spin healing $\xi_{\rm s}$ hence it is unlikely to have substantial effects.

\section{When is the Defect Density Determined?}\label{sec:beforeOrAfter}

Historically the KZ mechanism has been at the centre of a debate about the timing of the birth of defects \cite{bettencourt2000,Antunes:2006ty}. On one hand, it has been suggested that the appearance of defects can be traced to the fluctuations that freeze out at a time $\hat{t}$ \emph{before} the  transition occurs, i.e. when the dynamics first changes from being adiabatic to impulsive  \cite{Gill:1998ti}.  However,  there is also a symmetric point \emph{after} the transition, when the system dynamics again becomes adiabatic. An alternative viewpoint has been expressed that it is this point, \emph{after} the transition has occurred, when the eventual defect density is determined \cite{Antunes:2006ty,Deng:2009jo}. 

 In our numerical simulations we can probe the time at which the domain density is determined by employing a piecewise linear coupling quench for the ring-shaped system, with a quench time $\tau_Q$ that is different on each side of the critical point.  We choose a quench 
\begin{equation}\label{eq:asymmetric}
\hbar \Omega(t) = \left\{
					\begin{array}{ll}
						2\Omega_{\rm cr}\left(1-{2t}/{3 \tau_{\rm Q}}\right) & \mbox{if } \Omega(t)>\Omega_{\rm cr},\\
						2\Omega_{\rm cr}\left(1-{t/\tau_{\rm Q}}\right) & \mbox{if } \Omega(t)<\Omega_{\rm cr}.
					\end{array}
				  \right.
\end{equation}
For a given quench time $\tau_{\rm Q}$, from Eq.~\eqref{eq:asymmetric} we can see that the rate of change of $\Omega$ before the critical point is $3/2$ times smaller compared to the critical point. The choice of the relatively small factor of $3/2$ is due to the finite range of quenching times we can reliably simulate. Very fast quenches result in a larger number of domains that relaxes even after the end of the quench; on the other hand, very slow quenches are more likely to experience problems with the truncated Wigner approximation (Sec.~\ref{sec:uniform}). 

\begin{figure}
\begin{center}
\includegraphics[width=0.6\textwidth]{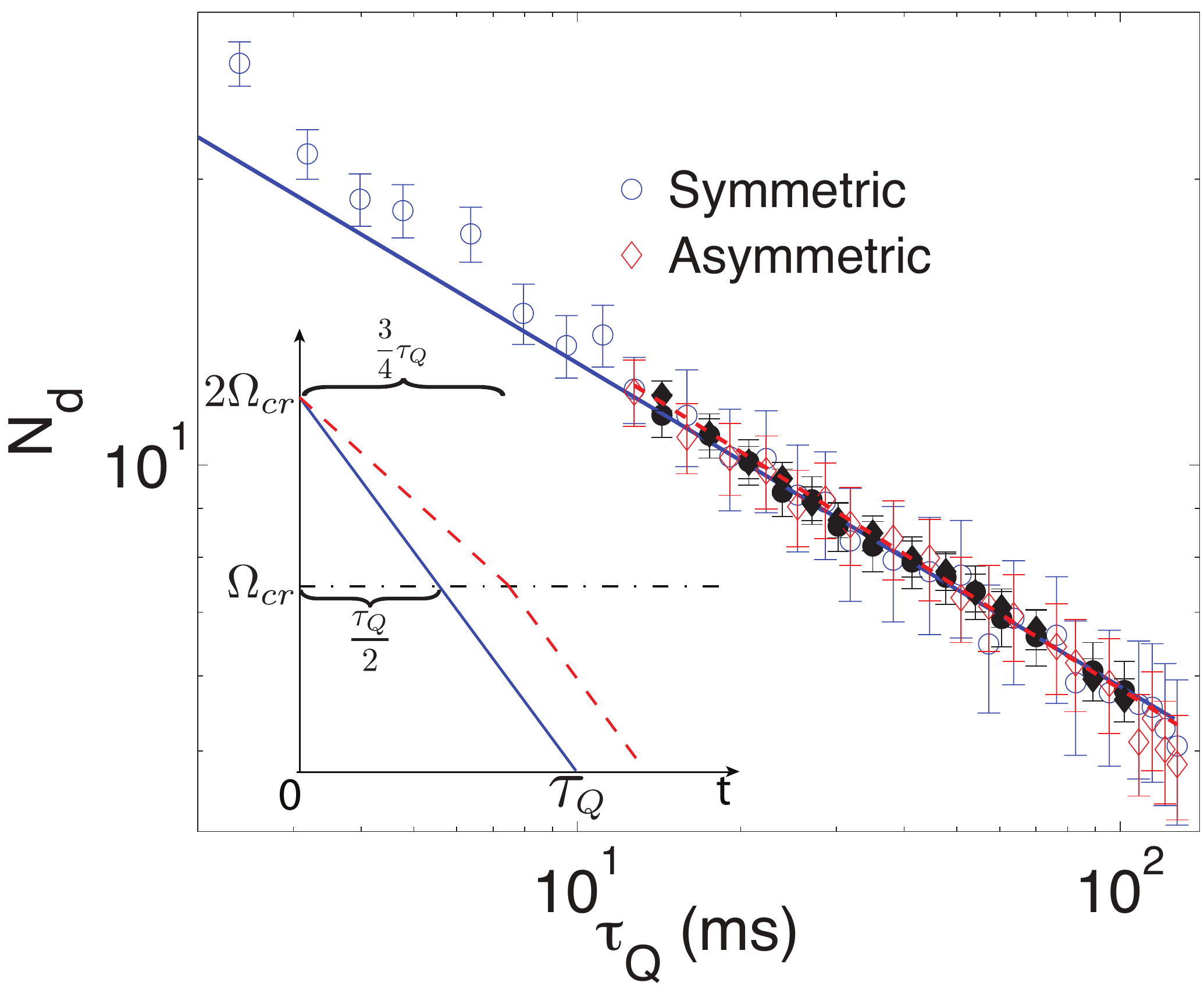}
\caption{Mean number of  domains formed for the asymmetric quench of Eq.~\eqref{eq:asymmetric} (open red diamonds) compared to that of the symmetric quench (open blue circles, black dots) as in Fig.~\ref{fig:uniform}(b). The two sets of data share the same quench time \emph{after} the critical point.  The agreement of the results suggests that the domains are imprinted \emph{after} the critical point.  The inset illustrates the difference in the coupling as a function of time.  Solid blue line: symmetric quench.  Dashed red line: asymmetric quench.}\label{fig:Asymmetric}
\end{center}
\end{figure}

We performed simulations of this system with the parameters otherwise as described in Sec.~\ref{sec:uniform}, and in Fig.~\ref{fig:Asymmetric} we plot the number of  domains formed as function of $\tau_{\rm Q}$ for this asymmetric quench. We fit the data with a power law $N_{\rm d} = a_{\rm asy}/\tau_{\rm Q}^{n}$ and derive the factor $a_{\rm asy}=8.74\pm 0.06$ which is in good agreement with the same quantity extracted from the data of Sec.~\ref{sec:uniform} in Fig.~\ref{fig:uniform}(c) of $a_{\rm uni}=8.63\pm 0.03$. 

The KZ mechanism does not provide a conclusive prediction for the total number of  topological defects formed in a phase transition, instead focusing on their scaling. This uncertainty is usually treated by saying that the typical distance between defects is $f \hat \xi$, where $f$ is unknown and typically varies from $1$--$10$ \cite{zurek1997}. However, in simulating the quenches in Fig.~\ref{fig:Asymmetric}, we have only altered the slope of the ramp \textit{before} the critical point, while all the other parameters remained identical. As a result it seems reasonable to assume that the pre-factor $f$ is the \textit{same} for both scenarios.

With the assumption that $f$ is fixed, if the defects were decided before the critical point, the absolute number of domains for the scenario with quench time $3/2$ times slower should have been reduced by a factor of $(3/2)^{n}=1.14$, where we have used $n=1/3$.  Then we would have expected to measure a prefactor of $a_{\rm asy}\approx7.54$.  This is well outside the uncertainty range of the value that we have measured.  The equivalence of the total number of domains for both type of quenches strongly suggests  the defect density is decided \emph{after} the transition.

This conclusion is compatible with the physical mechanism of domain formation in our system. In the miscible-immiscible quantum phase transition, domains form as a consequence of modulational instability of the system~\cite{Navarro:2009gd}. In this situation the population of unstable modes with complex eigenvalues grows exponentially \emph{after} the transition, suggesting that the number of formed domains is decided by the characteristic length of the most unstable mode at the moment of the return of adiabaticity.  

\section{Experimental Feasibility}\label{sec:feasibility}

We now examine the experimental feasibility of our scheme. The miscible and immiscible phases were experimentally observed in various binary systems \cite{McCarron:2011db} and recently both phases were realised with the same pair of atomic species using Feshbach resonances \cite{Papp:2008fg}. The results we present in this paper are  obtained in the regime where the two components are strongly immiscible with $\Delta = 0.25$.  This means that the system spin healing length is relatively short, and leads to both a large number of domains and their straightforward identification.  While such a strongly immiscible system may be difficult to realise in practice, it does not present an overwhelming obstacle to the experimental realisation of the scheme described here.

No pair of hyperfine states of $^{87}$Rb and $^{23}$Na naturally satisfy the criteria of naturally strong immiscibility, but the combination of the $|F=1,m_{\rm F}=+1\rangle$ and $|F=2,m_{\rm F}=-1\rangle$ hyperfine states in $^{87}$Rb has an interspecies Feshbach resonance \cite{Erhard:2004tb} that can be used to tune $\Delta$ to $0.8$ while keeping $g_{11}\approx g_{22}$. Although the use of a Feshbach resonance enhances losses, smaller values of $\Delta$ accelerate the process of domain formation allowing for the observation of the density pattern. We estimate that for a ring BEC with $L=50$ $\mu$m and $N=5000$, $\Delta=0.8$ yields $\xi_{\rm S} = 0.4$ $\mu$m, $\tau_{0}=351$ ms and $N_{\rm d}^{\rm max} \approx 50$.

Recently, Lin \emph{et al.} reported the observation of the miscible-immiscible quantum phase transition in spin-orbit coupled Bose-Einstein condensates  \cite{Lin:2011ba}. The group coupled two Zeeman sub-levels of the $F=1$ hyperfine state of $^{87}$Rb and was able to measure the phase separation of the dressed states across the critical point by changing the coupling strength $\Omega$. The method is not affected by increased inelastic losses, as outlined in the previous paragraph for Feshbach resonances, and is capable of achieving higher separation regimes ($\Delta \ll 1$) that make the task of counting  domains easier. Similar results have also been reported by Nicklas \emph{et al.}~\cite{Nicklas:2011el}. However, we note that the spatial configuration of the dressed state is not directly accessible, but is instead reconstructed from absorption imaging of the bare components. Hence, the higher separation and stability come at the price of a more complicated detection process for the number of domains. In addition, the results presented in Ref.~\cite{Nicklas:2011el} and our own numerical simulations suggest that the appearance of domains in the dressed state is slower than in the procedure proposed here. As a consequence, this alternative scheme could potentially result in higher domain loss before the counting, further altering the scaling exponent.

\section{Conclusions}\label{sec:conclusion}

In conclusion,  we have shown that the number of  defects formed in the coupling induced miscible-immiscible phase transition in a ring-shaped binary BEC scales as predicted by the Kibble-Zurek theory.  The scaling exponent we determine numerically for the number of defects for this system agrees with that predicted by using the mean-field critical exponents. The results suggest the scheme is a good candidate for the experimental testing of the predictions of the KZ mechanism in an experiment with ultra-cold gases, taking advantage of the system's isolation from the environment and the precise control of the parameter $\Omega$ that can be provided by microwave or laser coupling. 

We have also examined the phase transition in a binary BEC in an elongated harmonic trap, and found that this yields a modified scaling exponent compared to the homogeneous case. We have shown how an engineered coupling quench can be used to simulate inhomogeneous phase transitions in a ring BEC, and derived the scaling exponent of the harmonically trapped case independent of the domain counting problem inherent to trapped BEC systems. 

Engineered quenches can provide a systematic way to study the relatively unexplored problem of inhomogeneous phase transitions. Using this technique we have  shown how it is possible to invert the process and realise a homogeneous phase transition over a large spatial region of an harmonically trapped BEC.  Although we were unable to drive the entire condensate through the critical point at the same time due to the breakdown of our approximation, the approach clearly establishes a path to simplify phase transitions in inhomogeneous systems. 

Finally we have addressed two specific issues of the KZ mechanism: the relation between causality and defect formation, and the critical moment at which the defect density is decided. For the former we have derived a threshold value for the velocity of a front at which defect formation is suppressed in our system. For the latter, using a temporally asymmetric quench, we have supplied additional evidence in support of the case made in Ref.~\cite{Antunes:2006ty}, i.e. that  the density of defects is decided after the transition for this particular realisation of the KZ mechanism.

\ack
The authors thank Adolfo del Campo, Bogdan Damski, Markus Oberthaler, and Eike Nicklas for useful discussions.  This research was supported by the Australian Research Council through Discovery Project DP1094025, and the ARC Centre of Excellence for Engineered Quantum Systems. We acknowledge the support of U.S. Department of Energy through the LANL/LDRD program.

\appendix
\section{Critical Coupling and Critical Exponents}\label{app:criticalCoupling}

In this appendix we derive  an expression for the critical coupling $\Omega_{\rm cr}$ and critical exponent $\nu$ for a uniform binary BEC in a ring trap with $V(x) = 0$ (the definition of a uniform, homogeneous system). From Eq.~\eqref{eq:fullHam} follows that the time-independent Gross-Pitaevskii equation for $\psi_{i}$ is
\begin{equation}\label{eq:qGPE}
\mu\psi_{i}= \left[ -\frac{\hbar^2}{2m}\frac{\partial^2}{\partial x^2}+g_{ii}|\psi_{i}|^2+g_{12}|\psi_{3-i}|^2\right]\psi_{i} - \hbar\Omega \psi_{3-i}.
\end{equation}
Here we are interested in the behaviour of small perturbations around the mean-field solutions $\psi_{i}^{0}$ of \eqref{eq:qGPE}. We decompose the wavefunction $\psi_{i}$ into its mean field value $\psi_{i}^{0}$ and the fluctuations $\phi_{i}$ : $\psi_{i} = \psi_{i}^{0} + \phi_{i}$. Without loss of generality we take the condensate mean-field functions $\psi_{i}^{0}$ to be real. In order to simplify the equations further we work in the conditions where $g\equiv g_{11}=g_{22}\ne g_{12}$ and $N_{1}=N_{2}=N/2$. The equality of the intraspecies interactions is a good approximation for the reality of mixtures of hyperfine states of $^{87}$Rb, where the difference between the interaction constant is less than $5$\%. The equal number of particles  can easily be realised experimentally by starting with a single component, before applying a $\pi/2$ pulse of the coupling. If we substitute the decomposition of $\psi_{i}$ into Eq. \eqref{eq:qGPE} and retain only terms linear in $\phi$ we obtain the equation
\begin{equation}
\xi_{\rm B}^2\frac{\partial^2 \phi_{i}}{\partial x^2} = \sum_{j=1}^{2}S_{ij}\phi_{j},
\end{equation}
where $\xi_{\rm B} = \hbar/\sqrt{2mg\rho}$,
\begin{equation}\label{eq:linearisedGPE}
S = \left(\begin{array}{cc}
				2+\frac{\hbar\Omega}{g\rho} & 2\frac{g_{12}}{g}-\frac{\hbar\Omega}{g\rho}\\
				2\frac{g_{12}}{g}-\frac{\hbar\Omega}{g\rho} & 2+\frac{\hbar\Omega}{g\rho}
		  \end{array}\right)
\end{equation}
and $\rho = N/L$ is the linear density. The diagonalisation of the matrix $S$ results in a pair of eigenvalues $\Gamma_{i} = 2\{ 1 + 1/\sqrt{\Delta}, 1 - 1/\sqrt{\Delta} + \hbar\Omega/g\rho\}$ and a transformation matrix $U$ such that $A=USU^{-1}$ where $A$ is a diagonal matrix. In the basis $(\phi_{1}',\phi_{2}')^{T} = U (\phi_{1},\phi_{2})^{T}$, Eq.~\eqref{eq:linearisedGPE} takes the form
\begin{equation}\label{eq:diagGPE}
\frac{\xi_{\rm B}^2}{\Gamma_{i}}\frac{\partial^2 \phi_{i}'}{\partial x^2}=\phi_{i}'.
\end{equation}
We can then extract information about the miscible-immiscible phase transition by examining the effective correlation lengths $\xi_{i}=\sqrt{\xi_{\rm B}^2/\Gamma_{i}}$. As expected, an uncoupled mixture ($\Omega=0$) has a critical point for $\Delta=1$. We are interested here in the value of the critical coupling which turns an immiscible condensate into a miscible one. From $\xi_{2}$ it follows that when $\Delta<1$ the critical point is
\begin{equation}\label{eq:fixedCritical}
\hbar\Omega_{\rm cr} = g\rho(\frac{1}{\sqrt{\Delta}}-1).
\end{equation}
It is also possible to quantify the critical exponent $\nu$ by noticing that the effective correlation length scales as $\xi_{2} = \xi_{\rm B}/\Gamma_{2} = \xi_{\rm B}/|\epsilon|^{1/2}$, which combined with Eq.~\eqref{eq:corrLength} implies $\nu=1/2$.

Similar results for the critical coupling and critical exponent were obtained in Ref.~\cite{Lee:2009ib} from the energy spectrum of the system. We quote here only the energy gap $E_{\rm gap}\propto \hbar \sqrt{\Omega(t)[\Omega(t)-\Omega_{\rm cr}]}$ of the spectrum in the mean-field approximation. The presence of a gapped spectrum indicates that the phase transition is continuous. The relaxation time is related to the energy gap through the relation $\tau = \hbar/E_{\rm gap}\propto 1/|\epsilon|^{1/2}$ which implies the dynamical critical exponent is $z=1$.

\section*{References}
\bibliographystyle{unsrt}

\end{document}